\documentclass[11pt,a4paper]{article}

\usepackage[margin=1in]{geometry}

\usepackage[utf8]{inputenc}
\usepackage{url}
\usepackage{graphicx}
\usepackage{amsmath}
\usepackage{amssymb}
\usepackage{natbib}
\usepackage{hyperref}

\graphicspath{{figures_arxiv/}{figures/}}

\date{Submitted to JGR: Space Physics. Manuscript ID: 2026JA035326}

\begin{document}

\title{Multi-instrument constraints on a hemispherically asymmetric positive ionospheric storm in the 60--180$^\circ$E sector during the 12--13 November 2025 geomagnetic storm}

\author{
Pan Xiong$^{1}$,
Jianghe Chen$^{1}$,
Xuhui Shen$^{2,*}$,
Tong Liu$^{3,*}$,
Angelo De Santis$^{4}$,
Sergey Pulinets$^{5}$
\\[6pt]
\small $^{1}$Institute of Earthquake Forecasting, China Earthquake Administration, Beijing 100036, China\\
\small $^{2}$National Space Science Center, Chinese Academy of Sciences, Beijing, China\\
\small $^{3}$Department of Land Surveying and Geo-Informatics, The Hong Kong Polytechnic University, Hong Kong, China\\
\small $^{4}$Istituto Nazionale di Geofisica e Vulcanologia, Via di Vigna Murata 605, 00143 Rome, Italy\\
\small $^{5}$Space Research Institute, Russian Academy of Sciences, Profsoyuznaya str. 84/32, 117997 Moscow, Russia\\[4pt]
\small $^{*}$Corresponding authors: Xuhui Shen (shenxuhui@nssc.ac.cn), Tong Liu (tong2.liu@polyu.edu.hk)
}

\maketitle

\noindent\textbf{Key Points:}
\begin{itemize}
\item Multi-instrument observations reveal a positive ionospheric storm with clear hemispheric asymmetry in the 60--180$^\circ$E sector
\item F2-layer enhancement is density-dominated with no coherent sector-scale peak-height uplift, constraining mechanism interpretations
\item TEC/LSTID peaks precede strongest HF Doppler oscillations by several hours, indicating phase-dependent electrodynamic--neutral coupling
\end{itemize}

\begin{abstract}
Geomagnetic storms drive complex ionospheric responses through coupled electrodynamic and thermospheric processes, yet attributing storm-time TEC perturbations to specific mechanisms remains challenging. We investigate the ionospheric response to the 12--13 November 2025 intense geomagnetic storm (Dst minimum = $-$214~nT) in the 60--180$^\circ$E sector using a coordinated multi-instrument dataset comprising JPL GIM TEC, dense regional GNSS networks, continuous BeiDou GEO links, COSMIC-2 radio occultation, ground ionosondes, Swarm in-situ electron density, HF Doppler soundings, and TIMED/GUVI thermospheric composition observations. The observations reveal a dayside-dominant positive TEC storm with pronounced hemispheric asymmetry, where Northern Hemisphere mid-to-low latitudes exhibit stronger and longer-lasting enhancement than the Southern Hemisphere. Joint analysis of radio occultation, ionosonde, and Swarm data indicates that the enhancement is density-dominated with NmF2 and foF2 increases but with no coherent, sector-scale peak-height uplift in hmF2 or h$'$F2, posing challenges for uplift-only electrodynamic interpretations. Coherent large-scale traveling ionospheric disturbances propagate across the equator during UT~1--6, while HF Doppler oscillations maximize later during UT~6--24, revealing a timing offset between integrated TEC responses and reflection-height dynamics. Southern Hemisphere O/N$_2$ ratio depletion observed by TIMED/GUVI provides compositional context consistent with the faster positive-phase decay there, although concurrent Northern Hemisphere GUVI coverage is limited during this interval. These findings highlight the value of multi-observable diagnostics for developing testable constraints on storm-time mechanisms and improving sector-specific space weather nowcasting capabilities.
\end{abstract}

\section*{Plain Language Summary}
When an intense geomagnetic storm struck Earth on November 12--13, 2025, the ionosphere---the electrically charged upper atmosphere that affects GPS signals and radio communications---responded dramatically. Using multiple observation systems across the China--Australia region, we found that the ionosphere's electron content increased significantly during the storm's main phase, but this enhancement was notably stronger and lasted longer in the Northern Hemisphere than in the Southern Hemisphere. Our analysis indicates that this enhancement primarily reflected increased electron density near the ionosphere's peak layer rather than a persistent raising of the layer to higher altitudes, which challenges some commonly assumed mechanisms. We also discovered a timing mismatch where the strongest electron content increases occurred early in the storm while the most intense vertical oscillations happened several hours later, suggesting that different physical processes dominate at different storm phases. In the Southern Hemisphere, changes in the upper atmosphere's chemical composition may help explain why the enhancement faded faster there. These insights from combining multiple observation techniques improve our ability to understand and predict how space weather events affect navigation and communication systems in specific geographic regions.



\section{Introduction}
\label{sec:introduction}


Geomagnetic storms drive complex and multifaceted ionospheric variability that poses significant challenges for space weather specification and forecasting. Storm-time ionospheric responses manifest as large-scale total electron content (TEC) perturbations, steep density gradients, irregularities, and traveling ionospheric disturbances (TIDs), all of which can severely degrade Global Navigation Satellite System (GNSS) positioning accuracy and disrupt high-frequency (HF) radio communication \citep{buonsanto1999,prolss1995}. The equatorial ionization anomaly (EIA) is particularly susceptible to storm-time modulation through electric field penetration and thermospheric circulation changes, producing dramatic enhancements or depletions that propagate across local time and latitude \citep{balan2010,astafyeva2025_eia}. Intense geomagnetic storms represent especially valuable natural experiments for ionospheric physics because they simultaneously activate multiple coupled processes: prompt penetration electric fields (PPEFs) that rapidly redistribute low-latitude plasma \citep{fejer1979,tsurutani2004}, disturbance dynamo electric fields (DDEFs) driven by storm-enhanced thermospheric winds \citep{blanc1980,fejer1997}, large-scale traveling ionospheric disturbances (LSTIDs) launched by auroral heating \citep{hunsucker1982,hocke1996}, and thermospheric composition changes that modulate ion production and loss rates \citep{rishbeth1987,burns1991}. Understanding how these processes interact to produce observed storm-time signatures---and how those signatures differ between hemispheres, local times, and storm phases---remains a central objective of ionospheric research with direct implications for operational space weather services.


Despite decades of observational and modeling advances, several fundamental challenges persist in characterizing and attributing storm-time ionospheric responses. Positive ionospheric storms at low-to-mid latitudes have been extensively documented, with mechanisms including electric field-driven upward plasma drift, equatorward neutral winds that raise the F-layer into regions of reduced recombination, and modified production-loss balances \citep{prolss1995,fullerrowell1994,lu2008}. Negative storm phases, conversely, are typically attributed to storm-time compositional disturbances wherein molecular-rich air transported from high latitudes enhances recombination rates \citep{rishbeth1987,burns1991,fullerrowell1996}. However, distinguishing among these mechanisms using integrated TEC observations alone is problematic because different physical processes can produce similar signatures in column-integrated electron content \citep{buonsanto1999}. Specifically, a positive TEC enhancement may arise from F-layer uplift into lower-recombination regions, from genuine density increases at all heights, or from some combination---yet these scenarios have fundamentally different implications for the underlying physics and for predictions at other altitudes or locations.

A second challenge concerns the vertical structure of storm-time enhancements. Classic super-fountain scenarios invoke substantial F-layer peak height (hmF2) uplift as both an indicator and enabler of positive storms \citep{tsurutani2004,mannucci2005}, but observational constraints on height changes during individual events remain sparse relative to TEC measurements. Recent intense storms such as the May 2024 superstorm have demonstrated EIA expansion to unusually high latitudes \citep{astafyeva2024_superstorm}, yet the relationship between integrated TEC changes and vertical redistribution varies considerably between events. Without concurrent peak height and density information, mechanism interpretations often default to ``uplift-only'' narratives that may not universally apply.

Third, hemispheric asymmetries in storm-time ionospheric responses are commonly observed but often lack systematic multi-instrument characterization. Such asymmetries may arise from seasonal differences in background composition \citep{fullerrowell1996}, hemispheric differences in electric field mapping or conductivity \citep{hong2023,dang2016}, or simply from unequal observational coverage that obscures the true spatial structure of the disturbance. Recent comprehensive analyses of major storms \citep{barta2023,astafyeva2015} have underscored the importance of coordinated observations across both hemispheres to properly characterize and attribute these asymmetries.


The 12--13 November 2025 geomagnetic storm (Dst minimum = $-$214~nT) provides a well-instrumented test case to address these challenges within the 60--180$^\circ$E longitude sector spanning China, the West Pacific, and Australia. This event ranks among the more intense storms of the current solar cycle, featuring sustained southward interplanetary magnetic field (IMF Bz), enhanced solar wind dynamic pressure, and elevated auroral electrojet activity that persisted through the main and early recovery phases. The study sector offers several advantages for storm-time ionospheric investigation. First, the region hosts an unusually dense ground-based observing infrastructure, including the Crustal Movement Observation Network of China (CMONOC), the GNSS Earth Observation Network (GEONET) in Japan, the Geoscience Australia GNSS Archive (AGGA), the International GNSS Service (IGS), and multiple ionosonde stations---collectively providing comprehensive spatial coverage across both hemispheres \citep{schaer1998,hernandezpajares1999,reinisch2009}.

Second, and critically, the BeiDou Navigation Satellite System includes geostationary (GEO) satellites that maintain fixed positions over this longitude sector, enabling \textbf{continuous temporal monitoring} of ionospheric variability without the local-time drift inherent to medium-Earth-orbit (MEO) constellations \citep{liu2019_dcb,qian2019}. This geometric stability is uniquely suited for tracking the evolution of storm-time enhancements and hemispheric contrasts through the main and recovery phases.

Third, the availability of complementary space-based and ground-based measurements during this event---including Swarm in-situ electron density, radio occultation (RO) derived NmF2/hmF2 profiles from COSMIC-2 and other missions, HF Doppler soundings from the Meridian Project network, and thermospheric composition (O/N$_2$) observations from the Global Ultraviolet Imager (GUVI) aboard the Thermosphere, Ionosphere, Mesosphere Energetics and Dynamics (TIMED) satellite---permits multi-observable constraints that can distinguish among competing mechanisms and quantify the vertical structure of the storm response \citep{xiong2016,lei2007,lastovicka2006,zhang2004_guvi,meier2005}. The combination of global context from the JPL Global Ionospheric Map (GIM) TEC, sector-scale continuous monitoring from BeiDou GEO, and localized vertical structure information from multiple independent sources creates an observational framework rarely available for individual storm events.


In this study, we present a comprehensive multi-instrument analysis of the ionospheric response to the November 2025 storm in the 60--180$^\circ$E sector, articulating three contributions that address the challenges outlined above.

First, we characterize the global-to-sector TEC evolution in local-time coordinates and quantify the hemispheric asymmetry using continuous BeiDou GEO observations. We find that the Northern Hemisphere (15--50$^\circ$N) experienced stronger and longer-lasting positive storm enhancements---with peak relative slant TEC changes ($\Delta$STEC, defined by arc-start normalization) reaching 150--250~TECU and positive perturbations persisting for more than 12~hours---compared to Southern Hemisphere mid-latitudes, which decayed faster and exhibited earlier transition toward neutral or negative conditions. This asymmetry, documented with continuous sector-specific sampling unavailable from conventional MEO-only observations, provides observational constraints for hemispheric response studies and mechanism attribution.

Second, we constrain the vertical structure of the storm-time enhancement by jointly analyzing RO-derived NmF2/hmF2, ionosonde foF2/h$^\prime$F2, and Swarm topside electron density. Our observations indicate that the enhancement was primarily \textbf{density-dominated}, with significant increases in F2 peak and topside electron density, but with \textbf{no coherent, sector-scale peak-height uplift}. This constraint poses challenges for uplift-only interpretations and provides a testable benchmark for numerical model validation.

Third, we reveal coherent LSTIDs during the main phase with a dominant south-to-north propagation pattern across the equator, documented through GNSS-derived dTEC maps and latitude-time keograms \citep{tsugawa2004,ding2007}. Furthermore, we identify a pronounced \textbf{timing offset} between TEC/LSTID peak activity (UT~0--6) and strongest HF Doppler oscillations (UT~6--24), indicating that integrated electron content and reflection-region dynamics respond on different timescales to storm forcing. GUVI O/N$_2$ observations provide compositional context consistent with the faster Southern Hemisphere decay, although concurrent Northern Hemisphere GUVI coverage is limited during this interval \citep{crowley2006}. These findings illuminate the phase-dependent coupling among electrodynamic, neutral-dynamic, and compositional processes that together shape the sector-scale ionospheric response.


The remainder of this paper is organized as follows. Section~\ref{sec:data} summarizes the interplanetary drivers and geomagnetic context of the storm and describes the multi-instrument dataset employed for this analysis. Section~\ref{sec:methods} details the data processing methods, including GIM background definitions, GNSS-based dTEC/ROTI extraction, and quality control procedures for HF Doppler and other observations. Section~\ref{sec:results} presents results from global TEC evolution through sector-scale hemispheric asymmetry, traveling disturbances, and vertical structure constraints. Section~\ref{sec:discussion} discusses the testable mechanisms, comparisons with previous events, and observational uncertainties. Section~\ref{sec:conclusions} summarizes our findings and their implications for space weather diagnostics.


\section{Event Overview and Data Sets}
\label{sec:data}


\subsection{Interplanetary Drivers and Geomagnetic Response}
\label{sec:data:drivers}

\begin{figure}[htbp]
\centering
\includegraphics[width=\textwidth]{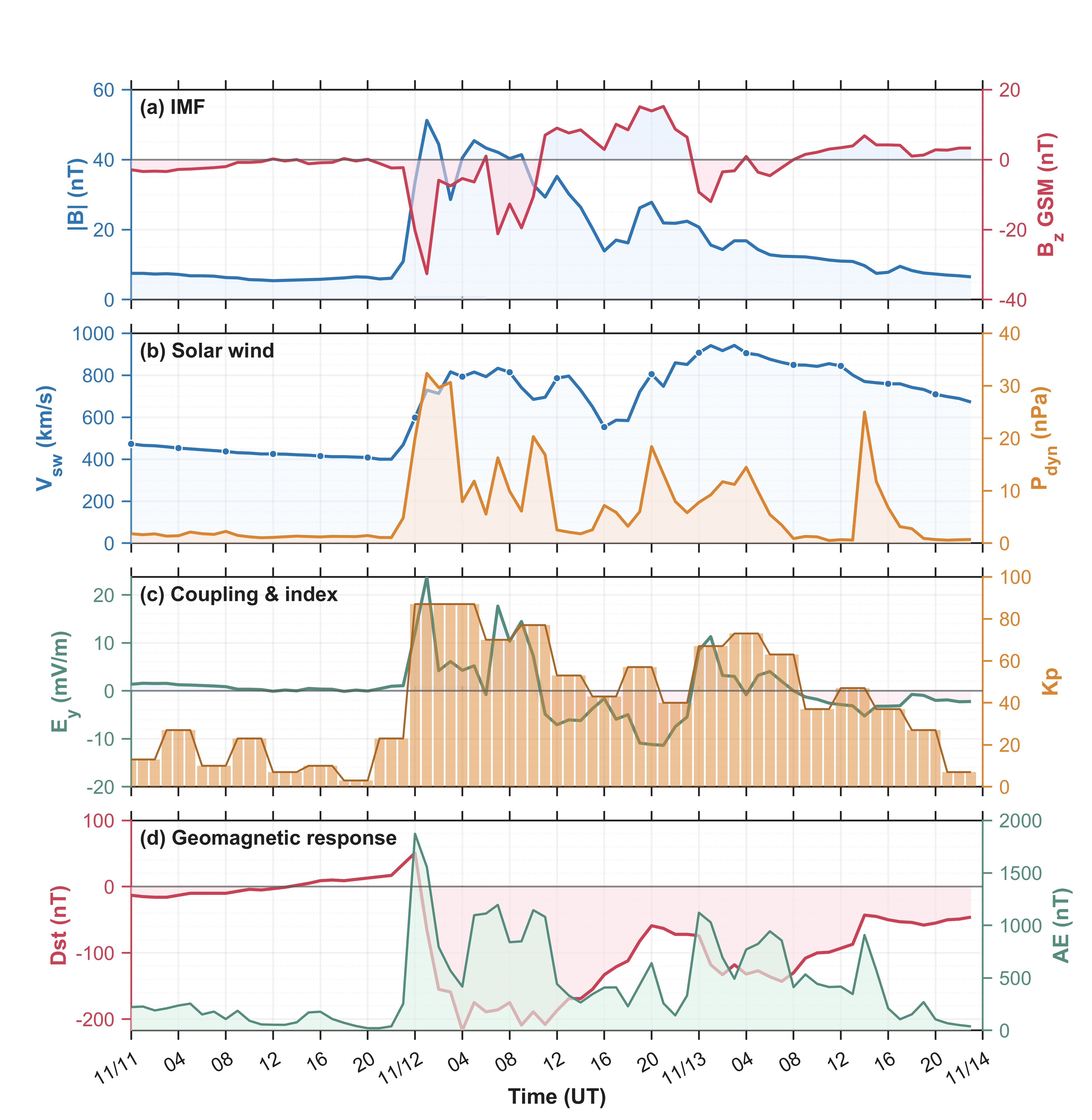}
\caption{Interplanetary driving conditions and geomagnetic response during 11--13 November 2025. (a) Interplanetary magnetic field (IMF) magnitude $|B|$ and $B_z$ component (GSM coordinates). (b) Solar wind speed $V_\mathrm{sw}$ and dynamic pressure $P_\mathrm{dyn}$. (c) Interplanetary dawn-dusk convection electric field $E_y = -V_x \times B_z$ (GSM). (d) Geomagnetic indices $Dst$ and $AE$. The shaded region indicates the storm main phase window (UT~0--6 on 12 November 2025).}
\label{fig:geomagnetic_indices}
\end{figure}

The 12--13 November 2025 geomagnetic storm was driven by a powerful interplanetary disturbance characterized by sharp increases in the interplanetary magnetic field (IMF) strength and sustained southward orientation. Figure~\ref{fig:geomagnetic_indices} presents the temporal evolution of the interplanetary and geomagnetic parameters during this event.

Beginning around 01--03~UT on 12 November, the IMF magnitude $|B|$ rapidly increased from quiet-time values of $\sim$5--8~nT to $\sim$40--50~nT, while the IMF $B_z$ component (in Geocentric Solar Magnetospheric, GSM, coordinates) underwent a pronounced southward turning, reaching minimum values near $-$30~nT and maintaining strongly southward conditions for an extended period (Figure~\ref{fig:geomagnetic_indices}a). Nearly simultaneously, the solar wind speed ($V_\mathrm{sw}$) increased sharply from $\sim$400~km~s$^{-1}$ to $\sim$750--850~km~s$^{-1}$, and the solar wind dynamic pressure ($P_\mathrm{dyn}$) exhibited multiple peaks exceeding 20~nPa, indicating substantial magnetospheric compression (Figure~\ref{fig:geomagnetic_indices}b).

Under the combined influence of strongly southward IMF and high-speed solar wind, solar wind--magnetosphere coupling was significantly enhanced. The interplanetary dawn-dusk convection electric field, calculated as $E_y = -V_x \times B_z$ (GSM), rapidly increased during the main phase, with peak values exceeding 15--20~mV~m$^{-1}$ (Figure~\ref{fig:geomagnetic_indices}c). Concurrently, the $K_p$ index rose sharply and remained at elevated levels, indicating widespread magnetospheric activity enhancement. Driven by this strong coupling, the $Dst$ index dropped precipitously between approximately 02--05~UT on 12 November, reaching a minimum of $-$214~nT, marking the main phase of an intense geomagnetic storm. Subsequently, $Dst$ entered a prolonged recovery phase characterized by multiple fluctuations associated with IMF reversals and variations in coupling strength (Figure~\ref{fig:geomagnetic_indices}d). Throughout the main and early recovery phases, the polar current systems exhibited marked intensification, with the auroral electrojet (AE) index sustained at levels ranging from hundreds to thousands of nT, indicating persistent high-latitude energy input.

For the ionospheric response analysis presented in subsequent sections, we define the storm main phase window as UT~0--6 on 12 November 2025, corresponding to the interval of strongest solar wind--magnetosphere--ionosphere coupling and most pronounced ionospheric disturbances within the study sector.


\subsection{Observational Assets in the 60--180$^\circ$E Sector}
\label{sec:data:observations}

\begin{figure}[htbp]
\centering
\includegraphics[width=\textwidth]{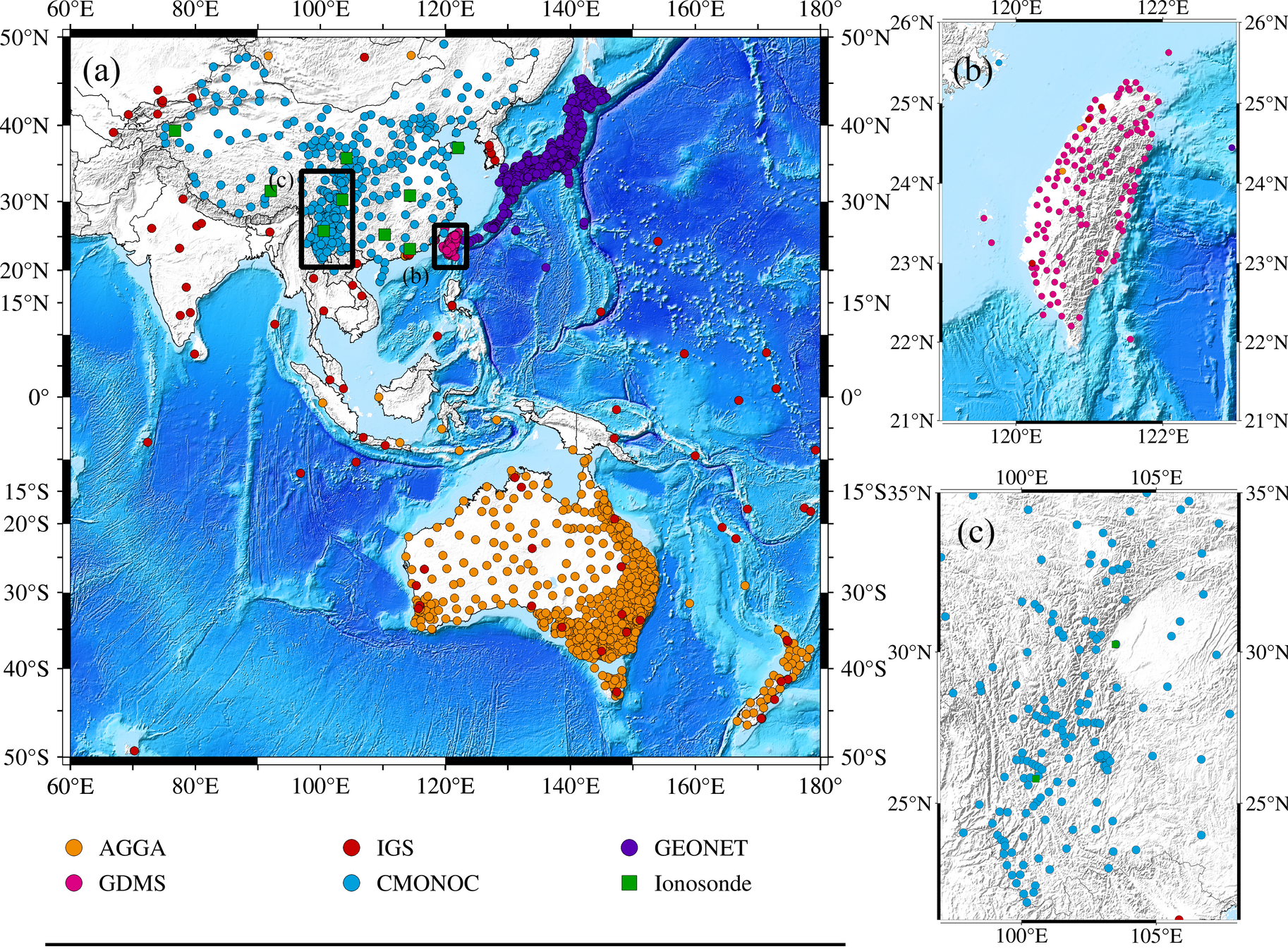}
\caption{Distribution of multi-source ionospheric observation networks in the China--Australia longitude sector (60--180$^\circ$E). Ground-based GNSS stations include CMONOC (China), IGS, GEONET (Japan), AGGA (Australia), and GDMS (Taiwan). Also shown are ionosonde stations and regional coverage areas.}
\label{fig:gnss_station_distribution}
\end{figure}

To characterize the ionospheric response during the 12--13 November 2025 storm, we employ a comprehensive suite of ground-based and space-based observations spanning the 60--180$^\circ$E longitude sector from China through the West Pacific to Australia. The spatial distribution of these observational assets is illustrated in Figure~\ref{fig:gnss_station_distribution}.

\subsubsection{Ground-Based GNSS Networks}
\label{sec:data:gnss}

Multiple GNSS receiver networks provide dense spatial sampling across the study sector for deriving TEC and its perturbations. In China and surrounding regions, we primarily utilize data from CMONOC GNSS stations, which offer high spatial density at low-to-mid latitudes enabling detailed characterization of TEC temporal evolution and latitudinal structure during the storm. Additional stations from IGS supplement the CMONOC coverage and extend observations beyond the Chinese mainland \citep{schaer1998,hernandezpajares1999}.

In Japan, GEONET provides high-density coverage for precisely resolving TEC spatial distributions in the northwestern Pacific mid-latitude region. In Australia, AGGA supplies observations from the Southern Hemisphere mid-latitudes, enabling cross-hemispheric comparative analysis. In the Taiwan region, the GPS Data Management System (GDMS) network further enhances sampling at the northern boundary of the equatorial ionization anomaly (EIA). This combined network configuration, spanning from $\sim$50$^\circ$N to $\sim$50$^\circ$S latitude, provides dense and spatially extensive coverage for investigating hemispheric asymmetries in the ionospheric storm response.

\subsubsection{BeiDou Geostationary Satellites}
\label{sec:data:beidou}

A distinctive advantage of the 60--180$^\circ$E sector is access to continuous ionospheric monitoring via the BeiDou Navigation Satellite System GEO satellites \citep{liu2019_dcb,qian2019}. Unlike medium-Earth-orbit (MEO) satellites whose ionospheric pierce points drift with orbital motion, BeiDou GEO satellites maintain fixed positions relative to the Earth, providing temporally continuous observations along stable ground-to-satellite paths. This geometry is particularly valuable for tracking storm-time TEC evolution and hemispheric contrasts without the confounding effects of local-time migration inherent to MEO-only observations.

\subsubsection{Space-Based Observations}
\label{sec:data:spacebased}

\paragraph{Swarm Satellite In-Situ Electron Density}
We employ in-situ electron density measurements from the European Space Agency (ESA) Swarm mission, specifically utilizing data from all three satellites (Swarm A, B, and C). These satellites operate in near-polar orbits at altitudes of approximately 450--550~km, providing calibrated topside ionosphere electron density from the Electric Field Instrument (EFI) Level~1B (L1B) data product \citep{xiong2016,paccagnella2020}. The electron density values are provided in SI units (m$^{-3}$) and require no additional calibration or scaling. Swarm observations are filtered to the UT~0--6 window and restricted to the study longitude and latitude ranges to ensure comparability across different days and satellites.

\paragraph{COSMIC-2 Radio Occultation}
To constrain the vertical structure of the ionospheric response, we utilize radio occultation (RO) data from the Constellation Observing System for Meteorology, Ionosphere, and Climate-2 (COSMIC-2) mission \citep{lei2007,cherniak2021_cosmic2,liu2022_cosmic2}. The RO technique provides vertical profiles of electron density from which F2-layer peak electron density (NmF2) and peak height (hmF2) can be retrieved. We collect RO profiles within the 60--180$^\circ$E longitude sector during the storm period and organize them by latitude and UT to examine the spatiotemporal distribution of F2-layer peak parameters. Level~2 (L2) ionospheric products are employed for this analysis.

\paragraph{TIMED/GUVI Thermospheric Composition}
To assess thermospheric composition changes potentially modulating the ionospheric response, we analyze observations from the Global Ultraviolet Imager (GUVI) aboard the Thermosphere, Ionosphere, Mesosphere Energetics and Dynamics (TIMED) satellite \citep{zhang2004_guvi,meier2005,crowley2006}. GUVI retrieves the thermospheric O/N$_2$ column density ratio from far-ultraviolet dayglow emissions, providing a proxy for the relative abundance of atomic oxygen (favorable for ionization) versus molecular nitrogen (enhancing recombination). We examine O/N$_2$ distributions over three consecutive days (DOY~315--317, corresponding to 11--13 November 2025) to identify storm-induced composition perturbations. We note that GUVI coverage of the Northern Hemisphere is limited during the selected observation periods due to orbital geometry constraints; consequently, the thermospheric composition analysis focuses primarily on the Southern Hemisphere (approximately 0--30$^\circ$S).

\subsubsection{Ground-Based Ionosondes}
\label{sec:data:ionosondes}

Multiple ionosonde stations within the Chinese sector provide ionogram-derived parameters including the F2-layer critical frequency (foF2) and virtual height (h$^\prime$F2) \citep{reinisch2009}. These ground-based measurements offer independent validation of RO-derived NmF2 and hmF2 and enable examination of local F2-layer evolution at specific geographic locations throughout the storm. The ionosonde observations complement the broader spatial coverage of GNSS and satellite-based techniques with high temporal resolution at fixed sites.

\subsubsection{HF Doppler Soundings}
\label{sec:data:hfdoppler}

To characterize ionospheric dynamical perturbations, we analyze HF Doppler observations from the Meridian Project network \citep{lastovicka2006,galushko2003,davies1966,Hao_2024_cmp_doppler}. HF Doppler measurements detect frequency shifts in reflected radio waves that arise from vertical motions of the ionospheric reflection layer, providing information complementary to the integrated electron content observed by GNSS. Multiple transmitter--receiver pairs within the study region enable spatiotemporal characterization of ionospheric vertical dynamics during the storm. We utilize the Meridian Project L2 data product, which provides pre-processed Doppler frequency shifts and derived velocities.

\subsubsection{Global Ionospheric Maps}
\label{sec:data:gim}

For global-scale context, we employ JPL GIM TEC products \citep{schaer1998,hernandezpajares1999}. JPL GIM products provide global TEC distributions with a temporal resolution of 1~h, enabling identification of storm-time positive and negative perturbations across different longitudes and local times. These global products establish the broader context within which the sector-specific analysis is situated.


\section{Methods}
\label{sec:methods}


\subsection{GIM TEC Background and Perturbation Definitions}
\label{sec:methods:gim}

To quantify storm-time ionospheric perturbations relative to undisturbed conditions, we construct a quiet-time background TEC reference from JPL GIM products. The quiet background is defined as the median TEC value computed over a sliding window of $\pm$13~days centered on the storm date (i.e., a 27-day window spanning approximately one solar rotation), following established practices for storm-time TEC perturbation characterization \citep{buonsanto1999,qian2019}. This median-based approach minimizes the influence of transient disturbances and provides a robust representation of typical ionospheric conditions.

The relative TEC perturbation is then calculated as the percentage deviation of the storm-time TEC from the quiet background:
\begin{equation}
\Delta\mathrm{TEC}\% = 100 \times \frac{\mathrm{TEC}_\mathrm{storm} - \mathrm{TEC}_\mathrm{quiet}}{\mathrm{TEC}_\mathrm{quiet}}
\label{eq:dtec_percent}
\end{equation}
where $\mathrm{TEC}_\mathrm{storm}$ is the observed TEC during the storm period and $\mathrm{TEC}_\mathrm{quiet}$ is the corresponding median background value at the same geographic location and universal time.

To suppress contamination from disturbed days mixed within the sliding window, a quartile-based (IQR) robust filtering is applied: for each grid point and each UT hour, the 25th and 75th percentiles ($Q_1$, $Q_3$) are computed along the day-of-year dimension, and the interquartile range is defined as $\mathrm{IQR} = Q_3 - Q_1$. Samples falling outside the interval $[Q_1 - 1.5 \times \mathrm{IQR},\, Q_3 + 1.5 \times \mathrm{IQR}]$ are flagged as outliers and excluded; the median of the remaining samples is then taken as the quiet-time background $\mathrm{TEC}_\mathrm{quiet}$. This method effectively preserves typical ionospheric states while reducing the influence of short-term strong disturbances or anomalous days on the background estimate.

This normalized metric facilitates comparison of perturbation magnitudes across regions with different background TEC levels and enables identification of positive (enhancement) and negative (depletion) ionospheric storm phases.


\subsection{GNSS-Based TEC Processing and Quality Control}
\label{sec:methods:gnsstec}

Ionospheric TEC values are derived from dual-frequency GNSS observations using established processing procedures \citep{hernandezpajares1999,pi1997}. The processing adopts an elevation angle cutoff of 15$^\circ$ to minimize multipath and mapping function errors at low elevations. The ionospheric pierce point (IPP) altitude is assumed to be 350~km, consistent with typical F-region peak heights and standard thin-shell ionospheric models.

Quality control procedures include cycle-slip and outlier detection based on the geometry-free (GF) combination and the Melbourne--W\"ubbena (MW) combination. Detection thresholds are set at 0.5~m for the GF combination and 5~m for the MW combination; epochs flagged as containing cycle slips or gross errors are rejected, and the affected observation arcs are re-segmented accordingly. This approach effectively prevents anomalous observations from contaminating subsequent TEC retrieval and perturbation extraction.

Observation arc segmentation follows these criteria: arcs are terminated and restarted when (1) the time gap between consecutive epochs exceeds 300~s, (2) the TEC jump between consecutive epochs exceeds 10~TECU, or (3) the satellite PRN changes. These conditions prevent pseudo-disturbances arising from data gaps, cycle slips, or satellite switches from being misinterpreted as geophysical signals.

Because the primary analysis focuses on relative TEC variations (e.g., dTEC and $\Delta$STEC) rather than absolute TEC values, hardware delay biases (differential code biases, DCBs) are not explicitly removed from the slant TEC (STEC) calculations. The dTEC extraction procedure described below inherently removes slowly varying biases through detrending, isolating the perturbation signals of interest; additional notes on relative TEC metrics are provided in Supporting Information Text~S1.


\subsection{dTEC and LSTID Extraction}
\label{sec:methods:dtec}

To extract traveling ionospheric disturbance (TID) signatures from GNSS TEC time series, we employ a ``detrending plus filtering'' approach consistent with recent studies of storm-time large-scale TIDs \citep{tsugawa2004,ding2007,ding2008,hunsucker1982,hocke1996}.

\subsubsection{Detrending via Savitzky--Golay Filtering}
\label{sec:methods:sg}

Within each continuous observation arc (segmented as described in Section~\ref{sec:methods:gnsstec}), a slowly varying background trend is constructed using Savitzky--Golay (SG) smoothing with a second-order polynomial and a window length of 121~samples. At the standard 30~s GNSS sampling interval, this window corresponds to approximately 1~hour, capturing background TEC variations associated with diurnal changes and satellite motion while preserving shorter-period fluctuations. The detrended perturbation sequence is defined as:
\begin{equation}
\mathrm{dTEC}(t) = \mathrm{STEC}(t) - \mathrm{STEC}_\mathrm{SG}(t)
\label{eq:dtec_sg}
\end{equation}
where $\mathrm{STEC}_\mathrm{SG}(t)$ is the SG-smoothed background trend. This detrending highlights medium-to-short period fluctuations relative to the slowly evolving background.

\subsubsection{Low-Pass Filtering for LSTID Extraction}
\label{sec:methods:butter}

To further suppress high-frequency noise and isolate LSTID signatures, a Butterworth low-pass filter is applied to the detrended dTEC sequences. The filter employs a 4th-order Butterworth low-pass design with a cutoff period of 40~min, retaining only perturbations with periods exceeding 40~min. Bidirectional zero-phase filtering (forward-backward, implemented via \texttt{filtfilt}) is applied to prevent phase distortion. These parameter choices are consistent with prior LSTID studies \citep{hunsucker1982,ding2007,ding2008} and effectively retain the dominant LSTID period range of approximately 40--180~min while attenuating medium-scale and high-frequency disturbances.

\subsubsection{Spatial Gridding and Keogram Construction}
\label{sec:methods:keogram}

To visualize the spatiotemporal evolution of LSTID wave fronts, the filtered dTEC values at individual IPP locations are spatially gridded onto a 0.3$^\circ$ $\times$ 0.3$^\circ$ (latitude $\times$ longitude) mesh. Within each grid cell, the mean dTEC from all contributing IPP observations is computed at each time step, reducing noise from multi-station and multi-satellite geometric variability.

Latitude--time keograms are then constructed by averaging the gridded dTEC values across the longitude range (60--180$^\circ$E), yielding a two-dimensional representation of perturbation amplitude as a function of latitude and time. In these keograms, coherent LSTID wave fronts appear as inclined stripes whose slope indicates the meridional propagation velocity and direction.


\subsection{ROTI Calculation}
\label{sec:methods:roti}

The Rate of TEC Index (ROTI) quantifies small-scale ionospheric irregularity activity through the temporal variability of TEC \citep{pi1997,cherniak2015}. In this study, ROTI is computed as the 5-min running standard deviation of the rate of change of VTEC (ROT), using 30-s GNSS sampling and requiring at least 50\% valid samples in each window. VTEC is obtained from STEC using a standard single-layer mapping function with a 350-km ionospheric shell height. Additional formulation details are provided in Supporting Information Text~S1.


\subsection{BeiDou GEO $\Delta$STEC Definition}
\label{sec:methods:beidou}

For the BeiDou GEO satellite observations, we focus on relative TEC variations rather than attempting absolute TEC calibration, because receiver and satellite hardware delays cannot be reliably separated on short timescales for GEO geometries \citep{liu2019_dcb}. To obtain a reproducible relative TEC change metric ($\Delta$STEC), we apply a start-point normalization procedure to each continuous station--satellite observation arc.

Let $\mathrm{STEC}(t)$ denote the slant TEC time series for a given arc. The reference starting epoch $t_\mathrm{ref}$ is selected as the second valid sample point within the arc (to mitigate edge effects at the arc boundary). The normalized relative TEC change is then defined as:
\begin{equation}
\Delta\mathrm{STEC}(t) = \mathrm{STEC}(t) - \mathrm{STEC}(t_\mathrm{ref})
\label{eq:dstec_beidou}
\end{equation}
This definition ensures $\Delta\mathrm{STEC}(t_\mathrm{ref}) = 0$ by construction, allowing inter-station and inter-latitude comparisons of the temporal morphology of TEC changes without the ambiguity introduced by unknown absolute biases. No hardware delay correction is applied; the analysis focuses solely on relative variations during the storm period.
Accordingly, $\Delta$STEC should be interpreted as a relative \textit{slant} TEC change from the arc-start reference epoch rather than an absolute TEC value (Supporting Information Text~S1).


\subsection{HF Doppler Data Processing}
\label{sec:methods:hfdoppler}

HF Doppler observations from the Meridian Project are analyzed to characterize ionospheric dynamical perturbations, particularly vertical motions of the F-region reflection layer \citep{davies1966,galushko2003,lastovicka2006}.

\subsubsection{Data Product and Variables}
\label{sec:methods:hf_product}

We utilize the Meridian Project Level~2 (L2) HF Doppler data product, which provides pre-computed values for each epoch including the Doppler frequency shift $\Delta f$ (Hz), equivalent Doppler velocity $V$ (m~s$^{-1}$), spectral width $W$ (Hz), and signal-to-noise ratio SNR (dB). The $\Delta f$ and $V$ values used in this study correspond to data columns 7 and 8 of the L2 files, respectively.

\subsubsection{Doppler Velocity Derivation}
\label{sec:methods:doppler_v}

The equivalent line-of-sight velocity is related to the Doppler frequency shift by:
\begin{equation}
V = \frac{c \times \Delta f}{2 \times f_0}
\label{eq:doppler_v}
\end{equation}
where $c$ is the speed of light and $f_0$ is the transmitter frequency (in MHz, provided in the L2 file header). This relationship captures the frequency shift arising from temporal changes in the ionospheric reflection path length. The transmitter--receiver link geometry (station coordinates) is also provided in the L2 file header; the analysis compares temporal evolution across different links without requiring precise reflection point reconstruction.

For the eight HF Doppler links analyzed in this study (Figure~\ref{fig:hf_doppler}), transmitter frequencies and station coordinates are summarized in Supporting Information Table~S1. For local-time context, we approximate a link local time as $\mathrm{LT} \approx \mathrm{UT} + \lambda/15$ (hours) using the link-midpoint longitude $\lambda$ (Table~S1), and note that the transmitter--receiver longitude span introduces only a small ($<1$~h) offset range across endpoints.

For physical interpretation, we adopt a nominal F-region reflection height range of approximately 200--300~km, consistent with typical HF propagation conditions.

\subsubsection{Quality Control}
\label{sec:methods:hf_qc}

Quality control criteria for the HF Doppler data include a signal-to-noise ratio (SNR) threshold of $\geq$10~dB, a spectral width limit of $W \leq 2$~Hz, and a requirement that QualityFlag = 1 where this flag is provided in the data product.
Additionally, robust outlier rejection is applied to the $V$ time series: samples deviating from the median by more than 6 times the median absolute deviation (MAD) are rejected, with corresponding $\Delta f$, $W$, and SNR values at the same epochs also removed.

For visualization purposes, a 2-min sliding median smoothing is applied to $\Delta f$ and $V$. However, all quantitative statistics (e.g., standard deviations computed separately for the UT~0--6 and UT~6--24 intervals) are based on the quality-controlled but unsmoothed data.


\subsection{RO, Swarm, Ionosonde, and GUVI Processing}
\label{sec:methods:other}

\subsubsection{Swarm Electron Density}
\label{sec:methods:swarm}

Swarm in-situ electron density observations utilize the calibrated electron density ($N_e$) from the EFI L1B data product for all three satellites (Swarm A, B, and C). No additional calibration or scaling is applied; the data are used directly in SI units (m$^{-3}$). For the comparative analysis across storm days, observations are filtered to the UT~0--6 window and restricted to the study longitude (60--180$^\circ$E) and latitude ranges to ensure consistent sampling conditions.

\subsubsection{Radio Occultation Gridding}
\label{sec:methods:ro}

COSMIC-2 RO profiles yielding NmF2 and hmF2 are organized into a latitude--time grid with cell dimensions of $\Delta t = 0.5$~h and $\Delta\mathrm{lat} = 5^\circ$. The number of RO profiles contributing to each grid cell is recorded to assess sampling density and statistical reliability. This gridding enables visualization of the spatiotemporal distribution of F2-layer peak parameters while accounting for the irregular sampling inherent to the RO technique.

\subsubsection{Ionosonde Data}
\label{sec:methods:ionosonde}

Ground-based ionosonde observations of foF2 and h$^\prime$F2 are used directly as time series without additional processing. These measurements provide independent local verification of the F2-layer peak density and height evolution observed by RO.

\subsubsection{GUVI O/N$_2$}
\label{sec:methods:guvi}

TIMED/GUVI O/N$_2$ column density ratio data are analyzed over three consecutive days (DOY~315--317, corresponding to 11--13 November 2025) to identify storm-induced thermospheric composition changes. Day-to-day comparisons highlight regions of O/N$_2$ depletion associated with storm-time neutral composition disturbances.

Given the limited Northern Hemisphere coverage during the selected observation periods (due to GUVI orbital geometry), the composition analysis focuses on the Southern Hemisphere, specifically the latitude range 0--30$^\circ$S within the study longitude sector. This regional focus constrains the thermospheric composition context for the observed hemispheric asymmetry in ionospheric response.


\section{Results}
\label{sec:results}

\subsection{Global TEC Evolution in Local-Time Coordinates}
\label{sec:results_global_tec}

Global JPL GIM TEC maps in local-time coordinates reveal a pronounced dayside-dominant positive storm during the main phase of 12 November 2025, providing essential context for the sector-focused analysis that follows \citep{buonsanto1999,prolss1995}. To characterize the storm-time ionospheric response relative to quiet conditions, TEC perturbations were computed as the percentage difference from a quiet reference defined as the median TEC over a $\pm$13-day (27-day) sliding window centered on the storm interval, following established GIM methodologies \citep{schaer1998,hernandezpajares1999}.

\begin{figure}[htbp]
\centering
\includegraphics[width=\textwidth]{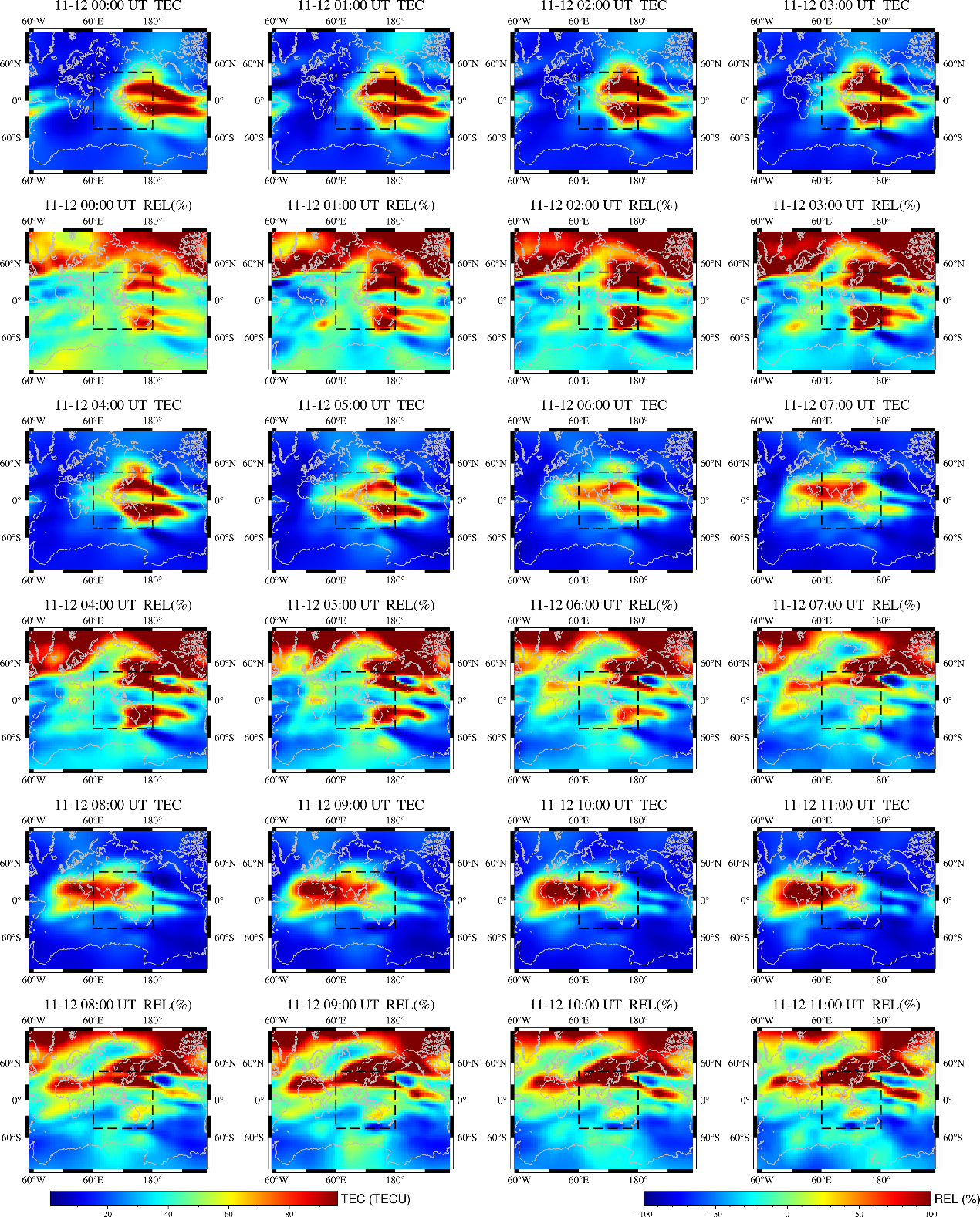}
\caption{\textbf{Global JPL GIM TEC and relative perturbation ($\Delta$TEC\%, based on 27-day sliding IQR background) during 00--11~UT on 12 November 2025.} For each UT snapshot, the upper panel shows TEC relative perturbation (\%) with respect to the IQR-filtered median background, and the lower panel shows absolute TEC (TECU). Maps are in local-time coordinates, highlighting the rapid establishment of a dayside-dominant positive storm during the main phase. For complementary global perturbation maps during the remaining periods (11 November 00--23~UT and 12 November 12--23~UT), see Supporting Information Figures~S1--S3.}
\label{fig:global_tec}
\end{figure}

During the pre-storm interval (11 November 23:00--12 November 00:00~UT), global TEC perturbations remained modest, with relative changes predominantly within $\pm$20\% and only scattered positive and negative perturbations at high latitudes and some mid-latitude regions (Figure~\ref{fig:global_tec}). The 60--180$^\circ$E sector (China--West Pacific--Australia) exhibited no systematic disturbance during this period, indicating that the ionosphere was close to its background state.

As the storm entered its main phase (approximately 12 November 01:00--06:00~UT), the global ionospheric response intensified rapidly (Figure~\ref{fig:global_tec}). Large-scale positive perturbations emerged across dayside low-to-mid latitudes, with relative TEC enhancements exceeding 50\% in multiple regions and locally approaching or exceeding 100\%, consistent with super-fountain effects reported during other intense storms \citep{mannucci2005,balan2010}. When viewed in local-time coordinates, these enhancements were strongly local-time dependent and were most pronounced on the dayside; at higher latitudes, substantial positive perturbations also extended into the nightside in some UT snapshots, whereas other nightside regions exhibited weak enhancement or negative perturbations. During this phase, the 60--180$^\circ$E sector was dominated by a positive ionospheric storm.

A notable feature during the early main phase was the appearance of a substantial positive storm response in the Southern Hemisphere Australian sector, which displayed an apparent equatorial conjugate relationship with the Northern Hemisphere China region \citep{dang2016}. This made Australia one of the few Southern Hemisphere mid-to-low latitude regions exhibiting a strong positive storm during this event. However, compared to the Northern Hemisphere, the Southern Hemisphere positive storm was shorter-lived, with its decay and disappearance occurring substantially earlier than the corresponding enhancement in China, revealing significant hemispheric asymmetry \citep{fullerrowell1994,hong2023}.

In terms of absolute TEC distribution, the equatorial ionization anomaly (EIA) structure strengthened markedly during the main phase, with a clear low-latitude double-crest morphology corresponding to the positive perturbation regions in the relative perturbation maps \citep{astafyeva2025_eia}. As the storm transitioned into the recovery phase (approximately 12 November 07:00~UT onward), the EIA structure within the 60--180$^\circ$E sector underwent pronounced adjustment (Figure~\ref{fig:global_tec}), with the original double-crest morphology gradually weakening and evolving toward a more single-crest configuration. Concurrently, a clear north-south hemispheric asymmetry emerged: Northern Hemisphere low-to-mid latitudes remained predominantly enhanced, whereas the Southern Hemisphere Australian sector experienced rapid decay of the positive storm and an earlier transition to weak or negative perturbation states. This asymmetry produced a complex spatial structure of coexisting positive and negative perturbations within the sector during the recovery phase.

To illustrate the full evolution of global perturbations throughout the storm, TEC relative perturbation maps for the remaining periods (11 November 00--23~UT and 12 November 12--23~UT) are provided in Supporting Information Figures~S1--S3.

\textit{Motivated by the strong response in 60--180$^\circ$E and its apparent hemispheric contrast, we next quantify the sector's latitude-dependent evolution using continuous BeiDou GEO links.}

\subsection{Sector TEC Evolution and Hemispheric Asymmetry from BeiDou GEO}
\label{sec:results_beidou}

\begin{figure}[htbp]
\centering
\includegraphics[width=0.9\textwidth]{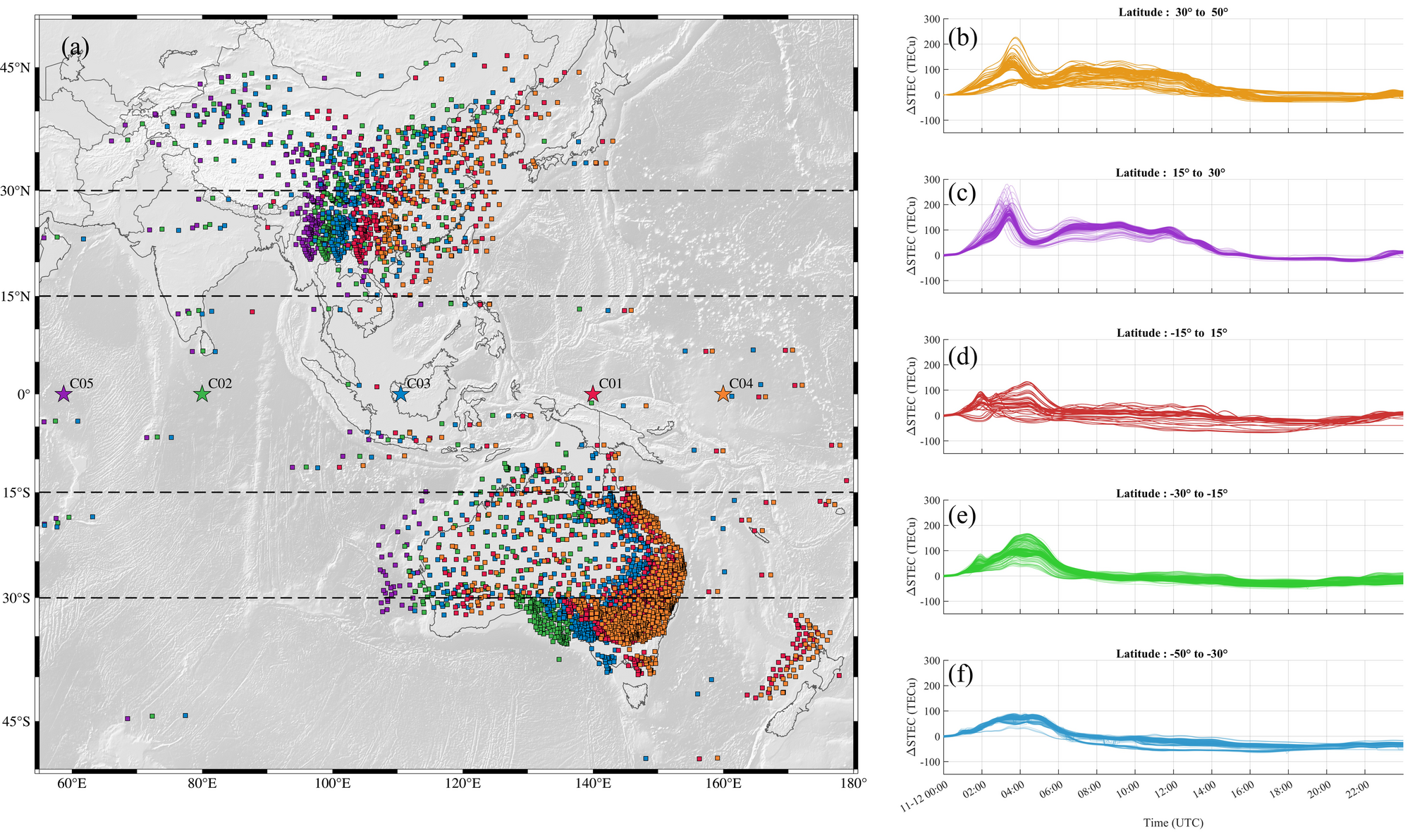}
\caption{\textbf{Sector-scale relative slant TEC change from BeiDou GEO observations.} (a) Ionospheric pierce point (IPP) distribution for BeiDou GEO satellite links in the 60--180$^\circ$E sector; dashed lines indicate the latitude bins used for statistics. (b--f) Time series of \textbf{$\Delta$STEC (TECU)} for five latitude bands (30--50$^\circ$N, 15--30$^\circ$N, $-$15--15$^\circ$, $-$30--$-$15$^\circ$, $-$50--$-$30$^\circ$) on 12 November 2025 (UT). $\Delta$STEC is defined as a relative slant TEC change from an arc-start reference epoch (no explicit hardware-delay calibration) and therefore should not be interpreted as an absolute TEC value. The GEO geometry yields long slant paths, so $\Delta$STEC magnitudes can exceed typical VTEC levels; here the emphasis is on timing, morphology, and inter-latitude/hemisphere contrasts.}
\label{fig:beidou_vtec}
\end{figure}

BeiDou GEO observations provide continuous sector-scale evidence that the positive storm was strongest in Northern mid-to-low latitudes and decayed substantially faster in Southern mid-latitudes \citep{liu2019_dcb,qian2019}. The geometric stability of GEO satellite-ground station links ensures consistent temporal sampling within a given longitude sector, making these observations particularly valuable for tracking storm-time TEC evolution. Since hardware delays are difficult to reliably separate on short timescales, absolute TEC calibration was not performed; instead, relative slant TEC changes ($\Delta$STEC) were computed using arc-start normalization to emphasize storm-time variations while enabling quantitative cross-latitude comparison (Figure~\ref{fig:beidou_vtec}).

Figure~\ref{fig:beidou_vtec}a shows the distribution of ionospheric pierce points (IPPs) for BeiDou GEO links within the study region, spanning from Northern Hemisphere mid-latitudes through the equatorial zone to Southern Hemisphere mid-latitudes, with dashed lines indicating the latitude bins used for statistical analysis. The $\Delta$STEC time series were binned into five latitude bands: 30--50$^\circ$N, 15--30$^\circ$N, $-$15--15$^\circ$, $-$30--$-$15$^\circ$, and $-$50--$-$30$^\circ$.

All latitude bands exhibited positive $\Delta$STEC enhancements during the early main phase (approximately 12 November 01:00--05:00~UT), but with significant differences in amplitude and duration. Northern Hemisphere mid-to-low latitudes (30--50$^\circ$N and 15--30$^\circ$N) displayed the most pronounced positive-phase perturbations, with peak $\Delta$STEC values generally reaching 150--250~TECU and showing good consistency among individual station-satellite links, reflecting a coherent positive storm response across this region. Notably, the $\Delta$STEC peaks in these two latitude bands occur during UT~02:00--04:00 (Figure~\ref{fig:beidou_vtec}b--c), earlier than the typical EIA diurnal maximum time in this longitude sector. The equatorial band ($\pm$15$^\circ$) exhibited weaker and smoother $\Delta$STEC enhancement compared to mid-latitudes, suggesting different response characteristics near the magnetic equator.

In contrast, Southern Hemisphere mid-latitudes ($-$30--$-$15$^\circ$ and $-$50--$-$30$^\circ$) showed positive $\Delta$STEC enhancements during the early main phase, but with notably smaller amplitudes (generally 50--150~TECU) and shorter durations compared to the Northern Hemisphere. As the storm progressed into the recovery phase, Southern Hemisphere $\Delta$STEC values returned more rapidly toward zero and in some cases became negative relative to the arc-start reference, while Northern Hemisphere mid-to-low latitude positive perturbations persisted for considerably longer. This pattern clearly indicates significant north-south hemispheric asymmetry in the storm-time ionospheric response: the Northern Hemisphere positive storm was stronger and longer-lasting, whereas the Southern Hemisphere positive storm was comparatively weaker and decayed more rapidly.

\textit{Given this rapidly evolving background and hemispheric asymmetry, we examine whether storm-time traveling disturbances and irregularities contribute to the sector response.}

\subsection{Storm-time Disturbances: dTEC Maps, ROTI, and Keogram Propagation}
\label{sec:results_disturbances}

GNSS-derived disturbance products show intense, coherent LSTID activity during UT~$\sim$1--6, with map-view wavefronts and a keogram indicating dominant south-to-north propagation across the equator, while irregularity activity is spatially localized \citep{hunsucker1982,zhang2023_lstid}.

\subsubsection{Map-View dTEC Evolution}

To characterize LSTID activity during the storm, detrended TEC (dTEC) was computed from regional GNSS networks using Savitzky-Golay background removal and Butterworth low-pass filtering (retaining periods $>$40~min). The dTEC values were gridded onto a 0.3$^\circ$$\times$0.3$^\circ$ (latitude$\times$longitude) mesh, with averaging applied to all IPPs within each grid cell to enhance signal visibility and suppress random noise.

Figure~\ref{fig:tid_maps} presents 20-minute cadence snapshots of dTEC spatial distribution from 01:00 to 06:00~UT on 12 November 2025; the complete time evolution is shown in Supporting Information Movie~S1. The most prominent LSTID activity occurred during this main-phase interval, characterized by organized, banded positive and negative structures (alternating colors) that indicate coherent traveling wavefronts rather than isolated station noise. Disturbance structures first appeared prominently over Southern mid-latitudes ($\sim$$-$50$^\circ$ to $-$20$^\circ$) early in the interval, with subsequent wave fronts extending northward across lower latitudes and into the Northern Hemisphere, suggesting a general south-to-north propagation pattern.

Over the East Asia--Northwest Pacific mid-latitudes, there were time windows where disturbance bands exhibited particularly sharp contrast, indicating strong coherence across many IPPs. Toward 05:00--06:00~UT, the disturbance field appeared less dominated by large, continuous bands than in earlier snapshots, suggesting a gradual weakening or fragmentation of the coherent LSTID activity.

\subsubsection{ROTI Distribution and Irregularity Context}

\begin{figure}[htbp]
\centering
\includegraphics[width=\textwidth]{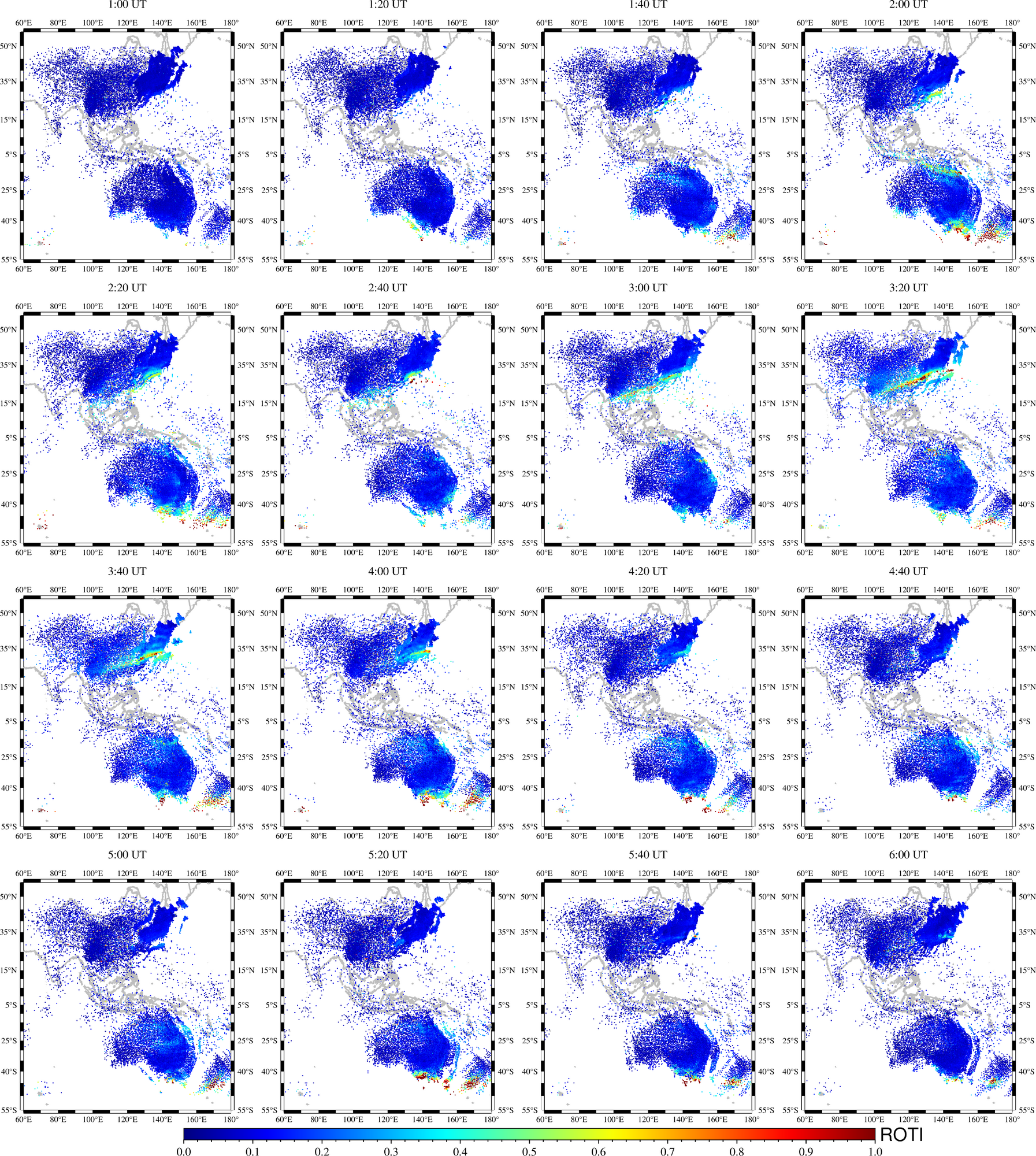}
\caption{\textbf{Spatiotemporal evolution of GNSS-derived ROTI in the 60--180$^\circ$E sector during UT~01:00--06:00 on 12 November 2025.} ROTI is computed as the 5-min running standard deviation of the rate of change of VTEC (ROT), using 30-s GNSS sampling and requiring at least 50\% valid samples in each window (see Methods). Panels (20-min cadence) show localized enhancements of irregularity activity contemporaneous with the storm main-phase disturbance interval.}
\label{fig:roti_maps}
\end{figure}

To distinguish wave-like LSTID signatures from irregularity-related fluctuations, the Rate of TEC Index (ROTI) was computed as the 5-minute running standard deviation of the rate of change of VTEC \citep{pi1997,cherniak2015} (Figure~\ref{fig:roti_maps}). During the UT~1--6 interval, most observation points exhibited low ROTI values (dark blue), indicating that the background ionosphere was relatively smooth. However, localized regions showed elevated ROTI (green/yellow/red), suggesting storm-time enhancement of irregularity activity that was spatially structured rather than uniform.

The timing of stronger ROTI patches overlapped portions of the interval when dTEC wavefronts were visually prominent, but the ROTI enhancements were more spatially localized than the broad dTEC wavefronts. This observation supports the interpretation that the keogram stripes in Figure~\ref{fig:keogram} represent coherent traveling disturbances rather than solely irregularity-driven fluctuations.

\subsubsection{Latitude-Time Keogram and Propagation Characteristics}

\begin{figure}[htbp]
\centering
\includegraphics[width=0.85\textwidth]{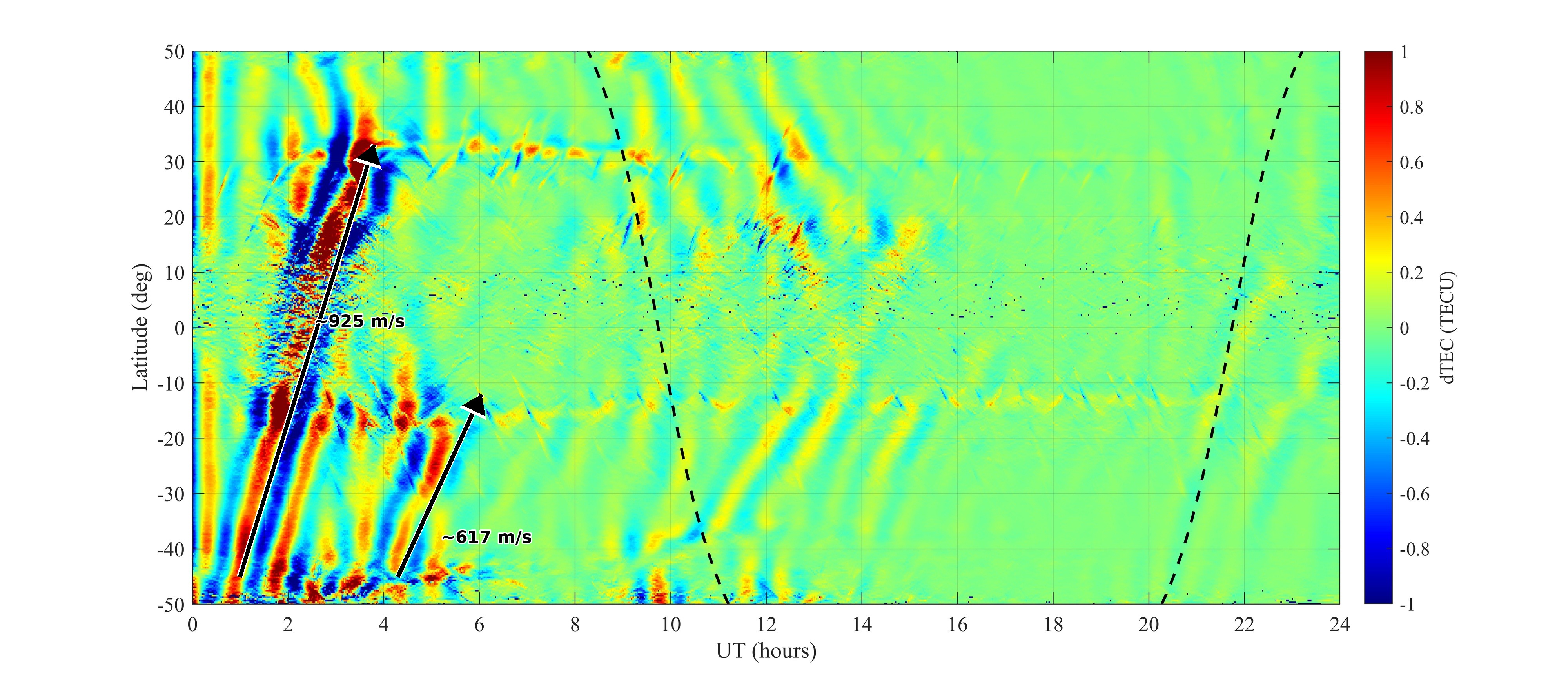}
\caption{\textbf{Latitude--time keogram of GNSS-derived TEC disturbance (dTEC) on 12 November 2025.} dTEC (TECU) is obtained by detrending GNSS TEC with a Savitzky-Golay filter (order 2, 121-sample window) and applying a 4th-order zero-phase Butterworth low-pass filter retaining periods $>$40~min. dTEC values are gridded on a 0.3$^\circ$$\times$0.3$^\circ$ (lat$\times$lon) mesh and averaged within each grid cell before constructing the latitude--UT keogram. The black dashed curves indicate the day--night terminator. Coherent diagonal bands during UT~$\sim$1--6 highlight storm-time LSTID activity with a dominant south-to-north (Southern-to-Northern Hemisphere) propagation signature; the stripe slopes correspond to meridional phase velocities of approximately 600--950~m/s based on a slope estimate from representative coherent bands (two annotated slope lines).}
\label{fig:keogram}
\end{figure}

To reveal the latitude-time evolution of LSTIDs and their propagation direction, a keogram was constructed by averaging the gridded dTEC over the 60--180$^\circ$E longitude range at each latitude and time (Figure~\ref{fig:keogram}). The most striking feature is the presence of strong, coherent diagonal banding (alternating positive and negative perturbations) during UT~$\sim$1--6, indicating a large-scale traveling disturbance with clear latitude-time progression.

The diagonal stripes generally tilt from earlier times at Southern mid-latitudes toward later times at Northern latitudes, consistent with a \textbf{dominant south-to-north propagation} across the equator. Here, ``south-to-north'' denotes propagation from Southern mid-latitudes toward the Northern Hemisphere (i.e., away from a likely Southern Hemisphere high-latitude source region), rather than a wave generated in the north propagating southward. At Southern mid-latitudes ($\sim$$-$50$^\circ$ to $-$20$^\circ$), disturbances appeared first with stronger amplitudes, suggesting this region was closer to the disturbance source or experienced earlier external forcing. The disturbances then propagated northward with stable inclination, crossing the equator and entering Northern mid-to-low latitudes (0--40$^\circ$N), maintaining good coherence across multiple latitude bands.

Near the equatorial region ($\pm$10$^\circ$), stripe contrast diminished somewhat, indicating reduced visibility or coherence. This attenuation likely reflects the strong background gradients associated with the EIA, which can mask disturbance signals of comparable amplitude rather than indicating propagation termination.

After UT~$\sim$8, the keogram showed weaker and more fragmented disturbance signatures, indicating reduced coherence and/or amplitude in later phases. The transition from ``strong, coherent propagating wave trains'' to ``weak, scattered residual disturbances'' coincided with the gradual recovery of regional TEC from its main-phase enhancement.

\textit{To interpret how these disturbances relate to the integrated TEC enhancement, we next constrain the storm-time vertical structure using independent peak and topside measurements.}

\begin{figure}[htbp]
\centering
\includegraphics[width=\textwidth]{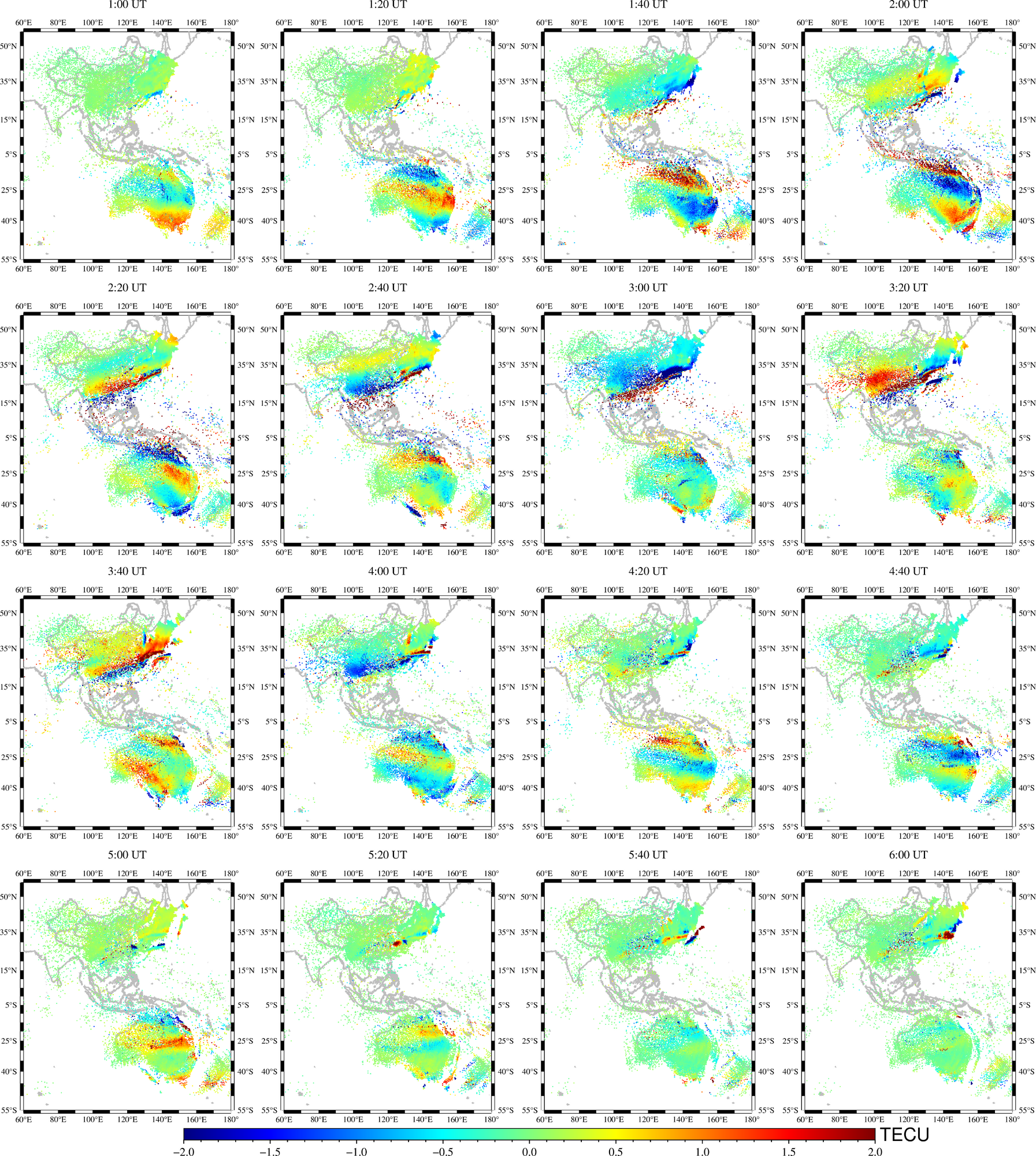}
\caption{\textbf{Map-view evolution of GNSS-derived TEC disturbance (dTEC) in the 60--180$^\circ$E sector during the storm main phase.} Panels show dTEC (TECU) at 20-min cadence from 01:00 to 06:00~UT on 12 November 2025. dTEC is obtained from GNSS TEC time series by arc segmentation, detrending with a Savitzky-Golay filter (order 2, 121-sample window), and a 4th-order zero-phase Butterworth low-pass filter retaining periods $>$40~min (see Methods). The sequence highlights coherent, traveling wave-like disturbance structures across the China--West Pacific--Australia region.}
\label{fig:tid_maps}
\end{figure}

\subsection{Vertical Structure Constraints: Swarm, RO, and Ionosonde Observations}
\label{sec:results_vertical}

Swarm in-situ Ne, RO-derived NmF2/hmF2, and ionosonde foF2/h$^\prime$F2 consistently indicate that the storm-time enhancement is density-dominated, with no coherent, sector-scale peak-height uplift \citep{xiong2016,paccagnella2020}.

\subsubsection{Swarm Topside Electron Density Response}

To examine the storm-time ionospheric response from a topside perspective, Swarm A/B/C EFI L1B electron density (Ne) data were analyzed for UT~0--6 within the 60--180$^\circ$E sector on three consecutive days: 11 November (pre-storm), 12 November (storm main phase), and 13 November (post-storm) (Figure~\ref{fig:swarm_ne}).

On 11 November (UT~0--6), the topside ionosphere exhibited relatively steady electron density with typical background distributions in the low-latitude region and no significant enhancement or anomalous structures at mid-to-high latitudes, establishing a reliable quiet-day baseline.

On the storm main-phase day (12 November, UT~0--6), Swarm satellites observed \textbf{pronounced electron density enhancement} in the mid-to-low latitude region within the China--Australia sector. The enhanced Ne regions showed continuous high-density values along satellite tracks, with amplitudes distinctly higher than the pre-storm day. Spatially, this enhancement corresponded well with the positive TEC storm region and the active LSTID zone identified from GNSS observations, demonstrating that the storm-time ionospheric disturbance was reflected not only in integrated TEC but also simultaneously in topside electron density structures at $\sim$450--550~km altitude. Notably, inter-track amplitude variations indicated spatial non-uniformity in the topside response rather than a homogeneous overall uplift.

On 13 November (UT~0--6), topside electron density had largely recovered, with most orbital segments showing Ne levels close to or below pre-storm background values, with only localized weak enhancement remnants. This indicates that the topside enhancement was strongly phase-dependent, consistent with the overall ``strongest during main phase, weakening during recovery'' evolution observed in TEC and LSTID analyses.

\subsubsection{Radio Occultation Constraints on F2-Layer Peak Structure}

Radio occultation (RO) retrievals of F2-layer peak height (hmF2) and peak electron density (NmF2) within the 60--180$^\circ$E sector are shown in Figure~\ref{fig:cosmic_ro} \citep{lei2007,yue2014}. The latitude-time scatter distribution reveals that during the main-phase interval (UT~0--6), \textbf{NmF2 exhibited significant enhancement} in the mid-to-low latitude region, with the enhancement band showing latitudinal continuity that corresponds well with the TEC positive storm and Swarm topside Ne enhancement zones.

In contrast, \textbf{hmF2 shows no coherent, sector-scale uplift} corresponding to the NmF2 enhancement. Most RO events show F2-layer peak heights distributed mainly in the $\sim$220--320~km range, with limited temporal variation. A small number of profiles exhibit transient height increases (e.g., a cluster reaching $\sim$350--400~km near $\sim$20$^\circ$S around UT~$\sim$04), but these are localized and not synchronized across latitude/time bins. This ``peak density significantly enhanced while peak height shows limited overall change'' feature indicates that the main-phase response was more prominently expressed as F2 peak electron density enhancement rather than a persistent layer-wide upward displacement.

To assess the statistical reliability of RO retrievals, Figure~\ref{fig:cosmic_ro}c shows the number of RO profiles per latitude-time grid cell ($\Delta$t = 0.5~h, $\Delta$lat = 5$^\circ$). Mid-to-low latitudes had continuous and adequate sampling coverage during the main phase and surrounding periods, confirming that the NmF2 enhancement and its latitudinal continuity revealed in Figures~\ref{fig:cosmic_ro}a--b have good statistical representativeness rather than being artifacts of individual profiles or uneven sampling.

\subsubsection{Ionosonde Validation of Local F2-Layer Response}

\begin{figure}[htbp]
\centering
\includegraphics[width=0.9\textwidth]{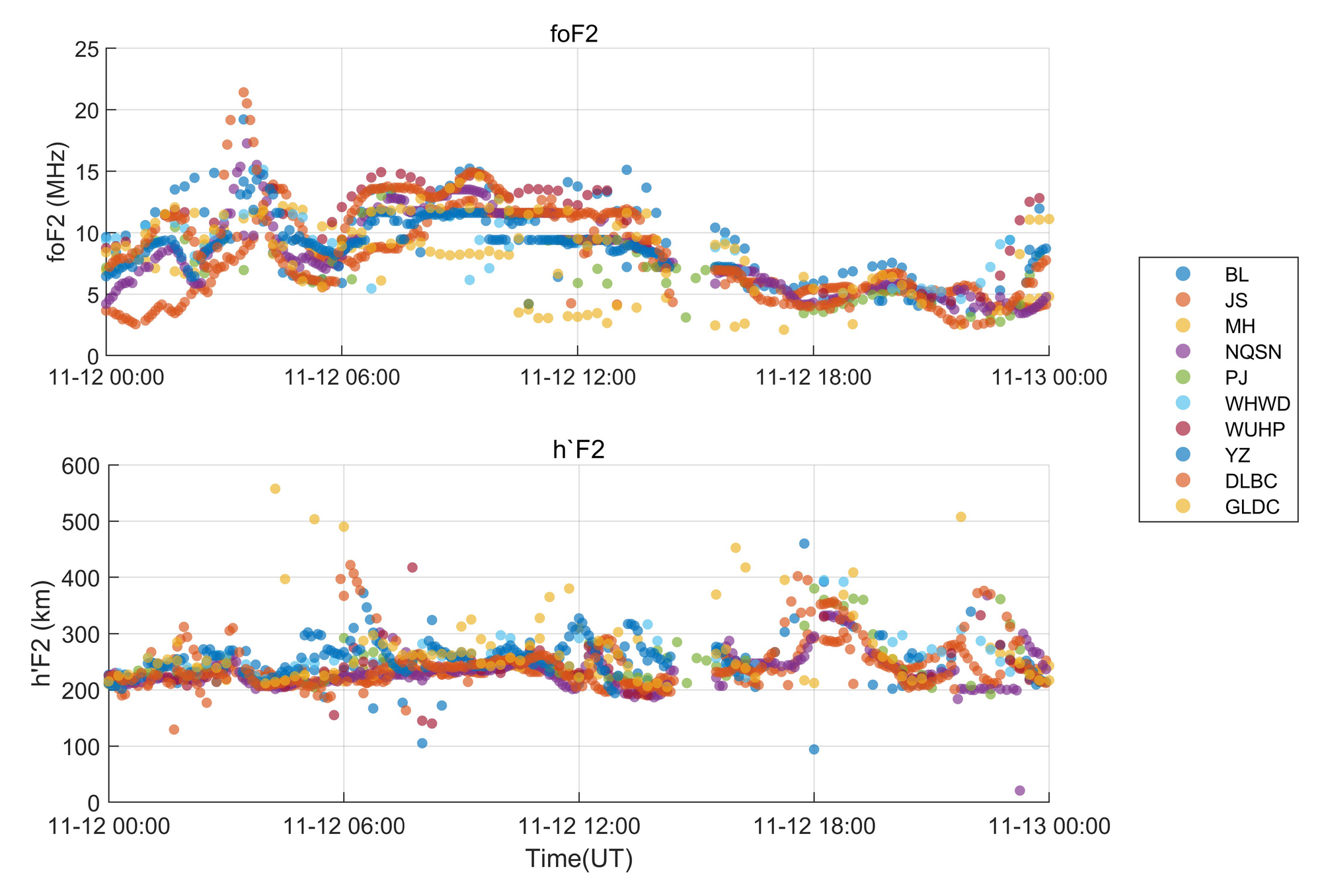}
\caption{\textbf{Ionosonde observations of F2-layer response in China during 12 November 2025.} (top) foF2 (MHz) and (bottom) h$^\prime$F2 (km) from multiple ionosonde stations (legend) plotted versus UT from 11-12 00:00 to 11-13 00:00. The foF2 enhancements during the storm main phase provide independent evidence for increased F2 peak electron density, while h$^\prime$F2 variations remain comparatively limited and non-systematic across stations.}
\label{fig:ionosonde}
\end{figure}

Ground-based ionosonde observations from multiple stations in China provide independent validation of the vertical structure response \citep{reinisch2009,buresova2014} (Figure~\ref{fig:ionosonde}). Multiple stations recorded notable \textbf{foF2 increases during the early main phase}, with peak values at some stations reaching 1.5--2 times their quiet-time levels, indicating significant F2-layer peak electron density enhancement at regional scales. This result shows good temporal and spatial consistency with GNSS TEC, RO-derived NmF2, and Swarm topside Ne enhancements.

In contrast to the pronounced foF2 changes, \textbf{h$^\prime$F2 (virtual height) did not exhibit sustained or consistent significant uplift} across most stations. Although individual stations showed transient height increases or scattered points at specific times, overall h$^\prime$F2 remained predominantly within the typical $\sim$200--300~km range, with variation amplitude notably smaller than the relative change in foF2. This indicates that at ground-based observation scales, the main storm-time F2-layer response was similarly manifested as peak density enhancement rather than systematic height elevation.

\subsubsection{Synthesis: Density-Dominated Enhancement Constraint}

Combining the independent observations from RO and ground-based ionosondes yields a consistent conclusion: during the main phase (UT~0--6), the ionosphere in the 60--180$^\circ$E sector exhibited \textbf{significant F2-layer peak electron density enhancement, while peak-height indicators (hmF2/h$^\prime$F2) did not show synchronized overall uplift} with the density changes. This vertical structure characteristic corroborates the TEC positive storm, BeiDou GEO $\Delta$STEC enhancement, and Swarm topside Ne increase results, demonstrating that the storm-time electron content increase was primarily achieved through density enhancement rather than simple layer-wide uplift.

This observational constraint provides a critical benchmark for mechanism interpretation: any explanation for the positive storm characteristics of this event must simultaneously account for ``significant peak density enhancement'' combined with ``limited overall peak height change.''

\textit{Finally, we assess storm-time ionospheric dynamics from HF Doppler measurements and place the hemispheric asymmetry in the context of Southern Hemisphere thermospheric composition changes.}

\begin{figure}[htbp]
\centering
\includegraphics[width=0.9\textwidth]{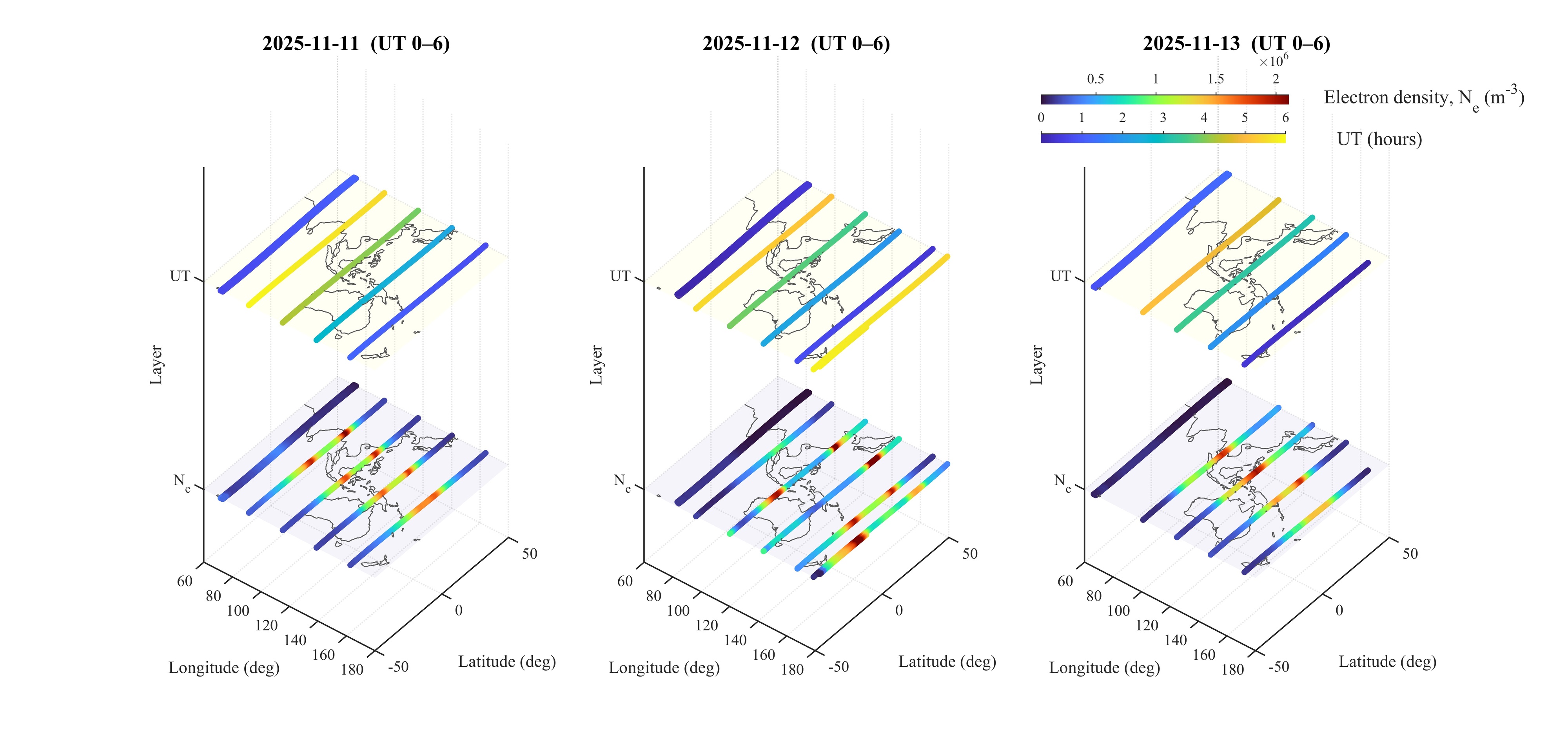}
\caption{\textbf{Swarm in-situ topside electron density during the storm interval (UT~0--6).} Swarm A/B/C EFI L1B electron density Ne (m$^{-3}$) is shown along satellite trajectories within 60--180$^\circ$E for (left) 11 November 2025 (pre-storm), (middle) 12 November 2025 (storm main phase), and (right) 13 November 2025 (post-storm), all for UT~0--6. Swarm samples the topside ionosphere at $\sim$450--550~km altitude (often near or above the F2 peak). The upper layer indicates the UT sampling along each trajectory, and the lower layer shows Ne along the same tracks, highlighting a pronounced topside density enhancement on the storm day.}
\label{fig:swarm_ne}
\end{figure}

\begin{figure}[htbp]
\centering
\includegraphics[width=0.9\textwidth]{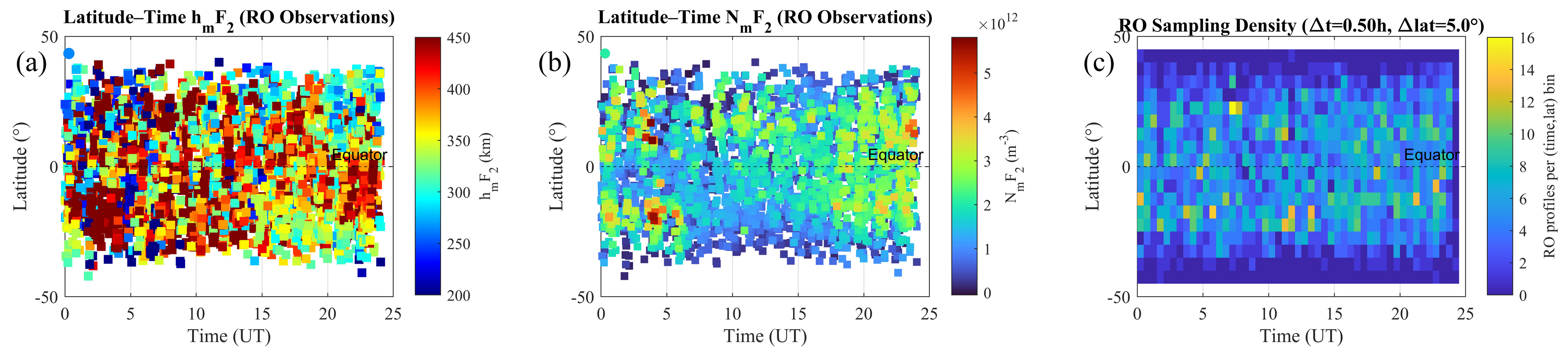}
\caption{\textbf{Radio occultation constraints on F2-layer peak structure in the 60--180$^\circ$E sector.} (a) Latitude--UT distribution of RO-derived hmF2 (km) and (b) NmF2 (m$^{-3}$) on 12 November 2025, using RO profiles within 60--180$^\circ$E. (c) RO sampling density (number of profiles per bin) for $\Delta$t = 0.5~h and $\Delta$lat = 5$^\circ$. The NmF2 enhancement during UT~0--6 provides a vertical-structure counterpart to the TEC positive storm, while hmF2 variations appear comparatively limited and non-uniform.}
\label{fig:cosmic_ro}
\end{figure}

\subsection{Reflection-region Dynamics and Thermospheric Composition Constraints}
\label{sec:results_dynamics_composition}

Multi-station HF Doppler records show that the strongest and most persistent oscillations occur in UT~6--24, later than the UT~0--6 TEC/LSTID peak, while GUVI observations indicate a storm-day reduction of Southern Hemisphere O/N$_2$ with delayed recovery \citep{lastovicka2006,zhang2004_guvi}.

\subsubsection{HF Doppler Dynamics and Timing Mismatch}

To characterize the storm-time ionospheric dynamical response from the perspective of reflection-region motions, HF Doppler observations from multiple stations were analyzed (Figure~\ref{fig:hf_doppler}). The Doppler frequency shift ($\Delta f$) primarily reflects line-of-sight motion information in the ionospheric reflection region (typically near the F-layer bottomside/peak vicinity), so rapid oscillations and positive-negative alternations indicate enhanced ionospheric vertical motion and modulation of the reflection height by propagating disturbances.

During the storm main phase (UT~0:00--6:00), most stations showed Doppler frequency shifts and equivalent velocities that were already enhanced relative to background levels, manifesting as ``disturbances beginning to appear, but continuous strong oscillations not yet at their most pronounced.'' Some stations exhibited intermittent oscillations, transient spikes, or weak quasi-periodic fluctuations during UT~0--6, indicating that main-phase energy input had already triggered motion responses in the ionospheric reflection region.

However, \textbf{more significant and sustained Doppler disturbances typically appeared after UT~6:00 and extended to 24:00}. Multiple stations (e.g., BOLU, WCWD, SZNS, SZZD, XGJL, CHYA) displayed stronger positive-negative swings, larger amplitudes, and longer-duration grouped oscillation structures during UT~6--24, indicating that ionospheric reflection-region vertical motion and wave processes were more intense and persistent during this period. The inter-station synchronicity of disturbance enhancement during this interval indicates this was not single-station noise or coincidental outliers, but rather a coherent regional-scale dynamical feature.

This ``Doppler-strongest-phase-shifted-later'' phenomenon forms an important contrast with the GNSS TEC/LSTID being strongest during the early main phase (UT~0--6): \textbf{UT~0--6 more prominently featured rapid establishment of TEC and large-scale wavefronts (dTEC keogram), while UT~6--24 more prominently featured intense fluctuations and oscillations of reflection-region heights}. In other words, TEC and Doppler responses are not necessarily synchronous in time: strong TEC changes can result from overall ionospheric ionization enhancement/transport, while Doppler is more sensitive to vertical motion and small-scale fluctuations near the reflection height, with its strongest phase potentially lagging behind the TEC peak phase.

\subsubsection{Southern Hemisphere Thermospheric Composition Constraint}

\begin{figure}[htbp]
\centering
\includegraphics[width=0.9\textwidth]{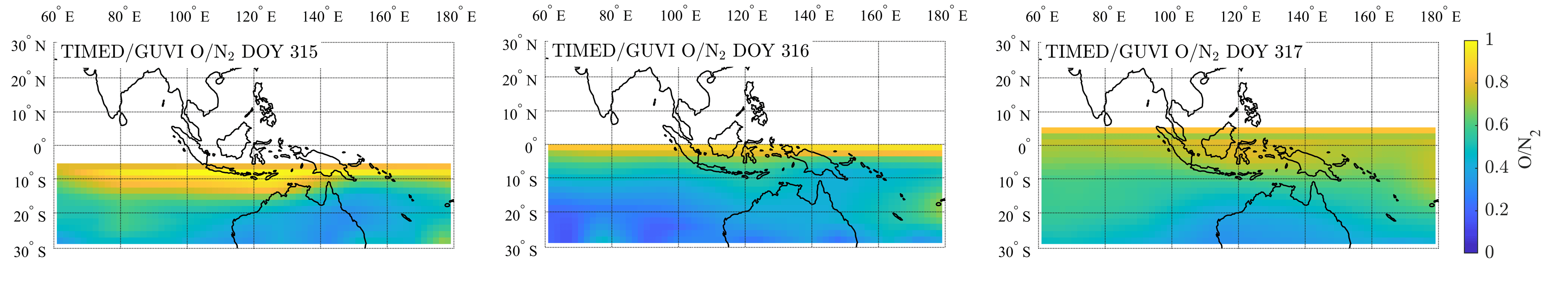}
\caption{\textbf{TIMED/GUVI thermospheric composition proxy (O/N$_2$) over the 60--180$^\circ$E sector.} Daily maps of GUVI-retrieved O/N$_2$ are shown for DOY~315 (11 Nov 2025), DOY~316 (12 Nov 2025), and DOY~317 (13 Nov 2025). Due to limited GUVI coverage in the Northern Hemisphere during this interval, interpretation is focused on the Southern low--mid latitudes (approximately 0--30$^\circ$S), where O/N$_2$ exhibits a storm-day decrease and incomplete recovery on the following day.}
\label{fig:guvi_on2}
\end{figure}

To constrain the thermospheric neutral composition conditions underlying the storm-time ionospheric response, TIMED/GUVI-derived O/N$_2$ ratio data were analyzed for three consecutive days (DOY~315--317) within the study sector \citep{meier2005,crowley2006} (Figure~\ref{fig:guvi_on2}). Due to limited GUVI coverage in the Northern Hemisphere during this observation window (with many pixels lacking data), the following analysis focuses primarily on the Southern Hemisphere mid-to-low latitudes (approximately 0--30$^\circ$S).

On the pre-storm day (DOY~315), Southern Hemisphere mid-to-low latitude O/N$_2$ exhibited a relatively smooth latitudinal distribution characteristic, with overall values at moderate levels representing a relatively quiet thermospheric composition state.

On the storm main-phase day (DOY~316), Southern Hemisphere mid-to-low latitude O/N$_2$ showed \textbf{pronounced reduction}, manifesting as a latitudinally-extended low-value band covering approximately 5--30$^\circ$S and spanning longitudinally across the China--Australia sector. Compared to pre-storm conditions, the O/N$_2$ reduction in this region was substantial, indicating enhanced molecular composition relative to atomic oxygen during the storm---a composition disturbance with strong spatial coherence in this region.

On the post-storm day (DOY~317), Southern Hemisphere mid-to-low latitude O/N$_2$ had partially recovered compared to DOY~316, but overall values remained lower than the pre-storm background level, with the low-value region still discernible. This indicates that the storm-induced composition disturbance in the Southern Hemisphere mid-to-low latitude region was characterized by prolonged duration and slow recovery.

\subsubsection{Linking Composition to Hemispheric Asymmetry}

The combined Southern Hemisphere observations indicate that thermospheric composition disturbance in the 60--180$^\circ$E sector during the storm was primarily manifested as a \textbf{significant decrease in Southern Hemisphere mid-to-low latitude O/N$_2$ with slow recovery}. This composition background characteristic provides an important constraint for interpreting the Southern Hemisphere ionospheric response: during the main-phase stage, although the Southern Hemisphere Australian sector also exhibited positive ionospheric enhancement conjugate with the Northern Hemisphere in the early phase, the lower O/N$_2$ conditions favor accelerated recombination processes, thereby limiting positive storm amplitude and promoting faster decay compared to the Northern Hemisphere.

Therefore, the TIMED/GUVI O/N$_2$ observations indicate that, at least in the observable Southern Hemisphere mid-to-low latitude region, storm-time thermospheric composition disturbance likely contributed to modulating the ionospheric response intensity and decay. Direct Northern Hemisphere comparison is not possible for this interval because GUVI coverage there is limited.

\begin{figure}[htbp]
\centering
\includegraphics[width=\textwidth]{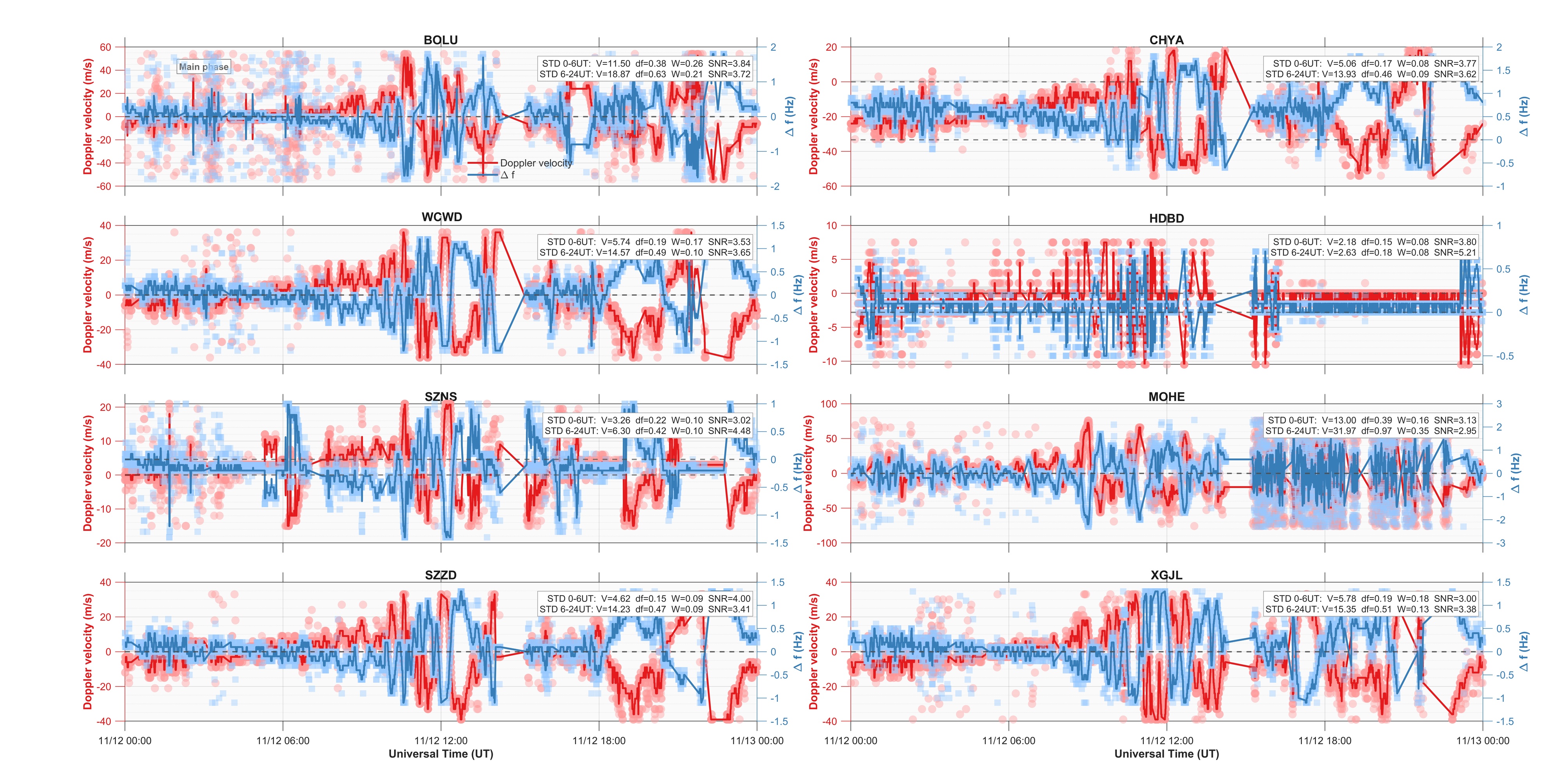}
\caption{\textbf{Multi-station HF Doppler observations of ionospheric dynamical response during 12 November 2025.} For each station link (titles), the Doppler frequency shift $\Delta f$ (blue, Hz; right axis) and equivalent Doppler velocity (red, m~s$^{-1}$; left axis) are shown versus UT from 11-12 00:00 to 11-13 00:00. Data are quality-controlled using SNR, spectral-width, and (if available) QualityFlag criteria, with robust outlier removal ($>$6$\times$MAD); 2-min running-median curves are shown for visualization. Inset boxes summarize disturbance strength (e.g., standard deviations) separately for UT~0--6 and UT~6--24, highlighting stronger, more persistent oscillations in the later interval.}
\label{fig:hf_doppler}
\end{figure}

\section{Discussion}
\label{sec:discussion}

The multi-instrument observations presented above provide a comprehensive view of the ionospheric response during the 12--13 November 2025 storm, revealing several features that merit physical interpretation and comparison with previous events. We organize this discussion along the key mechanistic pathways---electrodynamic forcing, traveling disturbance propagation, vertical structure constraints, hemispheric asymmetry, and observational considerations---highlighting four primary findings: (1) the density-dominated versus height-uplift constraint, (2) south-to-north LSTID propagation, (3) the TEC-Doppler timing mismatch, and (4) the compositional control on hemispheric asymmetry.

Although this work focuses on a single event, several results are most usefully interpreted as multi-observable constraints that can be tested in other storms and in numerical models. This framing is consistent with recent perspectives that emphasize increasingly integrated diagnosis of ionospheric variability across coupled processes, observations, and models \citep{radicella2026}. In particular, the combined density-versus-height evidence, the coherent early-phase traveling disturbances, and the timing offset between integrated TEC and reflection-region dynamics provide a set of observational relationships that any proposed mechanism should be able to reproduce. At the same time, the detailed spatiotemporal morphology and hemispheric contrast are expected to depend on storm forcing, background state, and local-time/longitude configuration, and are therefore likely to be partly event-specific.

\subsection{Electrodynamic Forcing and Rapid Positive Storm Establishment}
\label{sec:discuss_electrodynamic}

The rapid establishment of a dayside-dominant positive storm during the main phase (UT~0--6) is consistent with the classic prompt penetration electric field (PPEF) mechanism, whereby strong southward IMF drives magnetospheric convection that promptly penetrates to low latitudes, enhancing the eastward electric field on the dayside and intensifying the equatorial fountain effect \citep{fejer1979,tsurutani2004}. The observed TEC enhancements exceeding 50--100\% across low-to-mid latitudes (Figure~\ref{fig:global_tec}) and the strengthening of the EIA structure (Figure~\ref{fig:global_tec}) are hallmarks of PPEF-driven ionospheric uplift and poleward plasma transport \citep{mannucci2005,balan2010}.

Empirical models of storm-time electric fields indicate that the initial PPEF response can persist for 1--2 hours following southward IMF turnings \citep{fejer1997}, after which the disturbance dynamo electric field (DDEF) driven by storm-enhanced thermospheric winds becomes increasingly important \citep{blanc1980}. The temporal evolution of TEC enhancement in the 60--180$^\circ$E sector (peaking around UT~1--4 and gradually weakening thereafter) is qualitatively consistent with this PPEF-DDEF transition framework.

BeiDou GEO observations provide a complementary timing constraint on the sector-scale low-to-mid latitude response. In the Northern Hemisphere 30--50$^\circ$N and 15--30$^\circ$N bands, $\Delta$STEC exhibits a sharp peak during UT~02:00--04:00 (Figure~\ref{fig:beidou_vtec}b--c), earlier than the typical local-time of the EIA diurnal maximum in this longitude sector. This early peaking is consistent with rapid strengthening of the low-latitude eastward electric field during the early main phase, which can intensify the equatorial fountain effect and promote an earlier development of EIA crests. After UT~6, the subsequent evolution is more consistent with weakening penetration fields and a return toward background diurnal forcing.

In addition to the pronounced dayside enhancement, the global GIM perturbation maps show that substantial positive perturbations can extend into the nightside at higher latitudes during some UT snapshots (Figure~\ref{fig:global_tec}). This pattern may reflect storm-enhanced magnetospheric convection and associated plasma transport pathways that can redistribute high-density plasma from high latitudes toward mid-to-low latitudes under strong coupling conditions.

\textbf{A key observational constraint from our observations is that the positive storm was primarily density-dominated with no coherent, sector-scale peak-height uplift.} Radio occultation data show NmF2 enhancements of 1.5--2 times quiet values (Figure~\ref{fig:cosmic_ro}b), while hmF2 remained largely within the 220--320~km range without a synchronized upward displacement (Figure~\ref{fig:cosmic_ro}a). Ground-based ionosonde observations corroborate this pattern, with foF2 showing significant enhancement while h$^\prime$F2 variations remained comparatively limited and non-systematic (Figure~\ref{fig:ionosonde}). This does not exclude transient or localized uplift signatures, but it indicates that a persistent height rise is not the dominant sector-scale signature during the main phase. The resulting ``density up, height relatively unchanged'' constraint differs from the classic super-fountain picture in which both density enhancement and substantial layer uplift occur in tandem \citep{tsurutani2004}.

Comparison with previous intense storms provides additional context. During the October 2003 Halloween storms, \citet{mannucci2005} reported TEC enhancements exceeding 250\% with hmF2 uplifts of $>$100~km at some locations. Similarly, the May 2024 superstorm exhibited EIA expansion to unusually high latitudes accompanied by significant height changes \citep{astafyeva2025_eia,astafyeva2024_superstorm}. In contrast, the November 2025 event produced comparable positive-phase amplitudes but with notably weaker height response. This suggests that while electrodynamic forcing was sufficiently strong to drive substantial ionization enhancement, the vertical redistribution process---whether through modified meridional winds, competing compositional effects, or background state conditions---did not produce the pronounced layer uplift observed in other events.

The ``density-dominated with weak height response'' behavior highlights an important mechanistic puzzle for this event. One plausible interpretation is that PPEF-driven transport and fountain intensification rapidly enhanced low-latitude plasma densities, while other effects (e.g., evolving neutral winds, emerging disturbance dynamo electric fields, and storm-time composition changes that alter loss rates) acted to limit persistent, sector-scale hmF2 uplift at mid-to-low latitudes. This interpretation does not exclude transient or localized uplift signatures, but it emphasizes that the dominant sector-scale signature during the main phase is an electron density increase rather than a sustained layer-wide height increase.

Additional topside sounding evidence relevant to this vertical-structure interpretation is provided in Supporting Information (Text~S2 and Figures~S4--S5), which illustrates storm-time changes in the topside ionosphere at $\sim$820~km from the Ionosfera-M mission \citep{Pulinets_2026}.

\subsection{South-to-North LSTID Propagation and Neutral Atmospheric Disturbances}
\label{sec:discuss_lstid}

The keogram analysis (Figure~\ref{fig:keogram}) reveals a dominant south-to-north (Southern-to-Northern Hemisphere) propagation pattern for large-scale traveling ionospheric disturbances during UT~$\sim$1--6, with coherent diagonal banding extending from Southern mid-latitudes across the equator into Northern mid-latitudes. This propagation direction, combined with the timing that follows enhanced auroral electrojet activity during the main phase, is consistent with the generation of atmospheric gravity waves (AGWs) and traveling atmospheric disturbances (TADs) at high Southern latitudes through Joule and particle heating associated with auroral processes \citep{hunsucker1982,hocke1996}. In other words, the observed ``south-to-north'' signature is consistent with LSTIDs propagating away from a likely Southern Hemisphere high-latitude source region and crossing the equator under strong driving.

Statistical studies of storm-time LSTIDs in the Asian sector have established that periods greater than 40--60 minutes and phase velocities of 400--700~m/s are characteristic of equatorward-propagating disturbances originating from auroral latitudes \citep{tsugawa2004,ding2007}. The keogram stripe inclination in Figure~\ref{fig:keogram} suggests meridional phase velocities on the order of $\sim$600--950~m/s based on representative coherent bands (annotated slope lines), which lies near the upper end of typical storm-time LSTID speeds and should be interpreted as a rough, slope-based estimate given uncertainties in stripe selection and spatial averaging. The equatorial crossing of the disturbance wavefronts is also consistent with previous observations showing that LSTIDs can propagate across the magnetic equator under sufficiently strong driving conditions \citep{shiokawa2005,otsuka2004}.

The predominance of south-to-north propagation in this event may reflect the seasonal configuration. During mid-November (Southern Hemisphere late spring), the polar vortex structure and background thermospheric circulation favor efficient southward disturbance generation and subsequent northward propagation. Previous studies have noted that the seasonal distribution of LSTID propagation directions shows hemispheric dependence related to background wind patterns \citep{zhang2023_lstid}. In contrast, the October 2003 Halloween storms occurred closer to the autumnal equinox and exhibited more complex, multi-directional LSTID patterns.

The map-view dTEC snapshots (Figure~\ref{fig:tid_maps}) show organized wavefront structures with alternating positive and negative perturbations, while ROTI remained relatively low except in localized patches (Figure~\ref{fig:roti_maps}). This spatial pattern supports the interpretation that the dominant disturbance signature during UT~1--6 was from coherent traveling waves rather than irregularity-driven fluctuations, providing confidence in the keogram interpretation of propagating LSTIDs.

\subsection{Timing Offset Between TEC Enhancement and HF Doppler Oscillations}
\label{sec:discuss_timing}

\textbf{A notable finding is the temporal mismatch between the peak TEC/LSTID activity (UT~0--6) and the strongest HF Doppler oscillations (UT~6--24).} During the early main phase, integrated TEC showed rapid enhancement and LSTID wavefronts were most coherent, yet HF Doppler records displayed only moderate disturbance levels. Conversely, after UT~6, when TEC began its gradual recovery and LSTID coherence diminished, the HF Doppler signatures became substantially stronger and more persistent (Figure~\ref{fig:hf_doppler}).

Local-time context also differs between the two UT intervals for the HF Doppler links. The eight links are centered at longitudes 111.8--122.6$^\circ$E (Supporting Information Table~S1), corresponding to a local-time offset $\Delta$LT of 7.45--8.17~h ($\mathrm{LT} \approx \mathrm{UT} + \Delta\mathrm{LT}$). Thus, UT~0--6 corresponds to LT~$\sim$07:30--14:10 (local morning to early afternoon), whereas UT~6--24 spans LT~$\sim$13:30--08:10 (local afternoon through nighttime, wrapping across midnight). The later interval therefore increasingly samples post-sunset conditions, under which conductivity gradients and disturbance-dynamo-related electrodynamics may modulate reflection-region dynamics and favor stronger Doppler oscillations even as the integrated TEC response relaxes.

This timing offset can be understood by recognizing that TEC and HF Doppler sample different aspects of the ionospheric response. TEC represents the vertically integrated electron content and responds directly to large-scale ionization changes driven by electric fields, composition, and transport processes. In contrast, HF Doppler frequency shifts are primarily sensitive to vertical motions and refractive index variations near the reflection height (typically in the F-region bottomside to peak vicinity), making them useful tracers of wave-induced height fluctuations and smaller-scale perturbations \citep{davies1966,lastovicka2006}.

During UT~0--6, the dominant process was the rapid electrodynamic buildup of TEC enhancement, which would manifest as strong dTEC signals (as observed in the keogram) but may not necessarily produce large Doppler shifts if the electron density changes are relatively uniform in the vertical. After UT~6, as the driving electric fields weakened but neutral atmospheric disturbances continued to propagate and interact with the ionosphere, the sustained wave activity at reflection heights would generate persistent Doppler oscillations even as the overall TEC level began to recover \citep{galushko2003}.

This interpretation can be framed in a qualitative storm-phase timeline. During the early main phase, rapid energy input and prompt electrodynamic forcing can produce strong TEC changes through large-scale transport and ionization enhancement (PPEF-dominant conditions), whereas the subsequent interval can be increasingly shaped by neutral dynamics and delayed electrodynamic responses (e.g., disturbance dynamo), which can sustain wave activity and reflection-height oscillations even as the integrated TEC response relaxes. In addition, the dominant wave periods and vertical structures may differ between the early coherent LSTIDs (large-scale, regionally coherent signatures emphasized in TEC) and the later Doppler oscillations (potentially larger vertical velocity perturbations at the reflection height), which would further contribute to the observed timing offset.

This finding has implications for space weather monitoring. GNSS-based TEC observations and HF Doppler sounders provide complementary information: TEC captures the integrated ionospheric state and large-scale traveling disturbances, while HF Doppler is more sensitive to localized dynamics and wave activity at specific altitudes. Comprehensive characterization of storm-time ionospheric disturbances therefore benefits from multi-technique observations that capture both aspects.

\subsection{Hemispheric Asymmetry and Compositional Control on Southern Hemisphere Decay}
\label{sec:discuss_asymmetry}

\textbf{The pronounced hemispheric asymmetry, with the Northern Hemisphere positive storm stronger and longer-lasting than its Southern counterpart, is a key feature of this event.} BeiDou GEO observations show peak $\Delta$STEC of 150--250~TECU in Northern mid-to-low latitudes (30--50$^\circ$N) with positive perturbations persisting for $>$12~hours, whereas Southern mid-latitudes ($-$30 to $-$50$^\circ$S) exhibited peak values of only 50--150~TECU with the positive phase lasting approximately 4--6~hours before rapid decay (Figure~\ref{fig:beidou_vtec}).

In addition to the faster decay, a subset of Southern Hemisphere links exhibits $\Delta$STEC $<$ 0 during the later phase, indicating electron content below the arc-start reference. Because $\Delta$STEC is defined by arc-start normalization rather than a quiet-time baseline, such negative excursions do not necessarily imply that absolute TEC falls below pre-storm climatological conditions; they may also reflect a decline from a storm-enhanced arc-start level. Nevertheless, the global GIM-based $\Delta$TEC\% maps during the recovery interval (Supporting Information Figures~S1--S3) show that Southern mid-latitudes can transition toward weaker enhancement or locally negative perturbations, consistent with a rapid decay and possible depletion relative to quiet-time background in parts of the sector.

TIMED/GUVI observations provide compositional context for interpreting this asymmetry. The O/N$_2$ ratio in Southern mid-to-low latitudes showed significant reduction on the storm day (DOY~316; Figure~\ref{fig:guvi_on2}), indicating enhanced molecular composition (higher N$_2$ relative to O) that persisted into the recovery day (DOY~317) with incomplete restoration to pre-storm levels. Such O/N$_2$ depletion is consistent with accelerated ion-electron recombination through enhanced loss processes involving N$_2$ and O$_2$, which would tend to limit positive storm amplitude and promote faster decay \citep{rishbeth1987,burns1991}.

Theoretical and modeling studies have established that storm-enhanced thermospheric circulation transports molecular-rich air from high to low latitudes, and that this composition disturbance propagates equatorward with typical timescales of several hours \citep{fullerrowell1994,burns1991}. Seasonal asymmetry in background composition adds complexity: during November, the Southern Hemisphere is in late spring approaching summer, when background O/N$_2$ ratios are already lower than in the Northern Hemisphere winter \citep{fullerrowell1996}. This pre-existing hemispheric difference in composition may predispose the Southern Hemisphere to more rapid positive storm decay when additional storm-induced O/N$_2$ reduction occurs.

Comparison with other events supports the importance of composition. During the October 2003 Halloween storms, GUVI observations showed O/N$_2$ reductions exceeding 50\% in both hemispheres, with associated negative storm phases lasting several days \citep{crowley2006,meier2005}. The St. Patrick's Day 2015 storm also exhibited pronounced hemispheric asymmetry linked to composition changes \citep{astafyeva2015}. In the present event, the asymmetry is consistent with composition contributing to a faster Southern Hemisphere decay, with the Southern Hemisphere experiencing a ``double burden'' of storm-induced and seasonal O/N$_2$ depletion.

It should be noted that GUVI coverage in the Northern Hemisphere was limited during this observation window, precluding direct comparison of Northern versus Southern O/N$_2$ changes. The compositional interpretation for hemispheric asymmetry thus relies on Southern Hemisphere observations combined with theoretical expectations and seasonal climatology.

\medskip
\noindent\textit{Constraints versus ambiguities.}
The combined multi-instrument dataset provides several robust constraints, while leaving some attribution pathways only partially resolved:
\begin{itemize}
\item \textit{Constraints:} (i) a strong hemispheric contrast in the sector-scale relative TEC response from continuous BeiDou GEO links (Figure~\ref{fig:beidou_vtec}); (ii) a density-dominated enhancement with no coherent sector-scale uplift in hmF2/h$^\prime$F2 across independent instruments (Figures~\ref{fig:swarm_ne}--\ref{fig:ionosonde}); (iii) coherent early-phase LSTIDs with Southern-to-Northern Hemisphere propagation (Figures~\ref{fig:tid_maps}--\ref{fig:keogram}); and (iv) a timing offset between early TEC/LSTID activity and later HF Doppler oscillations (Figure~\ref{fig:hf_doppler}).
\item \textit{Ambiguities:} (i) limited Northern Hemisphere GUVI coverage prevents a direct interhemispheric comparison of storm-time composition changes; and (ii) the relative contributions of electric-field versus neutral-wind asymmetries to the hemispheric contrast cannot be quantitatively separated with the available composition and electrodynamic constraints.
\end{itemize}

\subsection{Unique Value of BeiDou GEO Observations and Observational Limitations}
\label{sec:discuss_limitations}

\textbf{BeiDou GEO satellite links provide unique advantages for storm-time ionospheric monitoring} that complement conventional GNSS observations. The geometric stability of GEO links---with satellites maintaining fixed positions relative to the rotating Earth---enables continuous temporal sampling of a given longitude sector without the diurnal local-time drift inherent to MEO constellation observations \citep{liu2019_dcb}. This stability is particularly valuable for tracking the evolution of storm-time enhancements and hemispheric asymmetry, as illustrated by the coherent latitude-banded patterns in Figure~\ref{fig:beidou_vtec}.

The arc-start normalization approach used for $\Delta$STEC computation emphasizes relative changes while avoiding absolute TEC calibration, which requires careful treatment of differential code biases (DCBs) that can be challenging to determine reliably for GEO satellites. This approach is well-suited for characterizing storm-time variations but does not provide absolute TEC values.

Several observational limitations should be acknowledged when interpreting the results presented. First, limited GUVI data availability in the Northern Hemisphere during this event precludes direct O/N$_2$ comparison between hemispheres, confining compositional interpretation primarily to Southern Hemisphere observations. Additionally, without explicit hardware delay removal, the BeiDou $\Delta$STEC values represent relative changes from arc start rather than absolute TEC, which may introduce baseline uncertainties for cross-station comparisons.

Methodological considerations also warrant attention. The Savitzky-Golay detrending and Butterworth low-pass filter parameters (window length, cutoff period) influence the extracted disturbance amplitudes and periods; while the 40-minute cutoff retains LSTID signatures, different parameter choices could emphasize different spatial scales. Furthermore, virtual height h$^\prime$F2 from ionograms differs from true hmF2 due to ray propagation effects, particularly under disturbed conditions with strong gradients, whereas the RO-derived hmF2 provides more direct peak height estimates but with sparse spatial sampling. COSMIC-2 and FY-3E GNOS provide variable sampling density across latitude and time, with some regions having limited profile coverage during specific storm phases. Finally, slope fitting to determine LSTID phase velocities from the latitude-time keogram involves statistical uncertainty that depends on the coherence and contrast of the diagonal stripes.

Despite these limitations, the multi-instrument consistency---with TEC, in-situ Ne, RO profiles, ionosondes, and HF Doppler all showing coherent storm-time responses---provides confidence in the findings. The combination of global (GIM), sector-scale (BeiDou GEO), and local (ionosonde, HF Doppler) observations enables cross-validation and mechanistic interpretation that would not be possible from any single data source alone.


\section{Conclusions}
\label{sec:conclusions}

Multi-instrument observations of the ionospheric response during the 12--13 November 2025 geomagnetic storm (Dst minimum = $-$214~nT) in the 60--180$^\circ$E sector yield the following conclusions:

\begin{enumerate}

\item \textbf{An intense geomagnetic storm with sustained magnetospheric coupling established a well-defined main-phase interval (UT~0--6) for coordinated ionospheric diagnostics} (Figure~\ref{fig:geomagnetic_indices}). The combination of persistent southward IMF, elevated interplanetary electric field, and strong auroral electrojet activity provided the external forcing that drove the ionospheric responses documented here. The event serves as a well-instrumented test case for storm-time coupling studies in a longitude sector with dense multi-network observational coverage.

\item \textbf{Global JPL GIM TEC maps reveal a dayside-dominant positive ionospheric storm with pronounced local-time dependence} (Figure~\ref{fig:global_tec}). The 60--180$^\circ$E sector exhibited strong positive TEC perturbations exceeding 50--100\% during the main phase, with the enhancement concentrated on the dayside. The subsequent emergence of hemispheric contrast highlights that sector-specific behavior cannot be inferred from globally averaged indices alone; local-time and longitudinal structure are essential considerations for space weather characterization.

\item \textbf{BeiDou GEO observations quantify significant hemispheric asymmetry in the storm-time TEC response} (Figure~\ref{fig:beidou_vtec}). Northern mid-to-low latitudes (15--50$^\circ$N) displayed peak $\Delta$STEC of 150--250~TECU with positive perturbations persisting for $>$12~hours, whereas Southern mid-latitudes ($-$30 to $-$50$^\circ$S) showed weaker enhancements (50--150~TECU) with the positive phase lasting only 4--6~hours before rapid decay. Here $\Delta$STEC represents a relative \textit{slant} TEC change from an arc-start reference epoch (not an absolute TEC value). The continuous temporal sampling afforded by GEO geometry provides unique advantages for tracking storm-time evolution and interhemispheric contrasts.

\item \textbf{GNSS-derived disturbance products reveal strong, coherent LSTIDs during UT~1--6 with a dominant south-to-north (Southern-to-Northern Hemisphere) propagation signature} (Figures~\ref{fig:tid_maps}, \ref{fig:roti_maps}, \ref{fig:keogram}). The keogram analysis shows organized wavefronts crossing the equator from Southern mid-latitudes into the Northern Hemisphere, consistent with gravity wave generation at high Southern latitudes during the main phase. The spatial localization of elevated ROTI indicates that irregularity activity, while enhanced, does not account for the broad coherent LSTID signatures, confirming that neutral atmospheric disturbances contribute a structured propagating component to the storm-time ionospheric response.

\item \textbf{Independent vertical-structure observations from Swarm, radio occultation, and ionosondes consistently indicate that the storm-time enhancement is density-dominated, with no coherent sector-scale peak-height uplift} (Figures~\ref{fig:swarm_ne}, \ref{fig:cosmic_ro}, \ref{fig:ionosonde}). Topside Ne, F2-peak density (NmF2/foF2), and integrated TEC all show substantial enhancement during the main phase, whereas hmF2 and h$^\prime$F2 remained largely within typical ranges (220--320~km) without synchronized upward displacement. Transient or localized height increases may occur but are not the dominant sector-scale signature. This ``density up, height relatively unchanged'' constraint implies that mechanisms invoking sustained layer-wide uplift alone are insufficient; models must account for density enhancement through processes that do not require persistent large-scale height increases.

\item \textbf{A timing offset exists between the strongest TEC/LSTID activity (UT~0--6) and the most pronounced HF Doppler oscillations (UT~6--24), and Southern Hemisphere O/N$_2$ depletion provides compositional context for the faster decay there} (Figures~\ref{fig:hf_doppler}, \ref{fig:guvi_on2}). The phase-dependent dominance of different observables indicates that integrated electron content responds rapidly to electrodynamic forcing during the main phase, while reflection-region dynamics become most active later as neutral atmospheric perturbations continue to interact with the ionosphere. GUVI observations of storm-day O/N$_2$ reduction in Southern mid-to-low latitudes, with incomplete recovery the following day, are consistent with a compositional contribution to the hemispheric contrast in storm-time TEC decay; direct Northern Hemisphere composition comparison is limited by GUVI coverage during this interval.

\end{enumerate}

Taken together, these observations indicate that the November 2025 storm produced a rapid, dayside positive TEC storm in the 60--180$^\circ$E sector that is strongly latitude-structured and hemispherically asymmetric. The enhancement is supported by both topside and F2-peak density increases while peak-height proxies show no coherent sector-scale uplift, providing observational constraints that any proposed mechanism must satisfy. Coherent LSTIDs dominate the early main phase with Southern-to-Northern Hemisphere propagation, whereas reflection-region dynamics inferred from HF Doppler peak later, indicating that different coupled processes can dominate different storm phases. The Southern Hemisphere O/N$_2$ depletion provides compositional context consistent with the interhemispheric contrast in positive storm persistence, while acknowledging limited Northern Hemisphere GUVI coverage for direct comparison.

While this study is event-based, the density-versus-height constraint and the multi-observable timing relationships provide testable benchmarks for model evaluation and for future comparative analyses across intense storms and background states.

These findings underscore the value of coordinated multi-instrument observations---spanning integrated TEC, in-situ topside density, F2-peak parameters, traveling disturbances, reflection-region dynamics, and thermospheric composition---for characterizing storm-time ionospheric responses and constraining the underlying mechanisms. The dense observational coverage in the 60--180$^\circ$E sector, including continuous BeiDou GEO monitoring, provides a template for sector-specific space weather diagnostics that can inform operational applications in navigation and communication systems.

\section*{Acknowledgments}
We thank the data providers for making their datasets publicly available.
This work was supported by the Natural Science Foundation of China (NSFC)
under grant numbers 42274108, U25A20563, and U2539201, the Central
Public-interest Scientific Institution Basal Research Fund
(No. CEAIEF2025030105), and the ESA-MOST Dragon 6 cooperation
(project ID 0095456 and 0095407).
Additional support was provided by the Space It Up project funded by the Italian Space Agency, ASI, and the Ministry of University and Research, MUR, under contract n. 2024-5-E.0 - CUP n. I53D24000060005.

\section*{Data Availability Statement}

The CMONOC-derived ionospheric electron content data are archived at Zenodo \citep{xiong_cmonoc_2026}. GNSS data were provided by IGS (\url{https://cddis.nasa.gov/gnss/}), GEONET (\url{https://www.gsi.go.jp/ENGLISH/}), AGGA (\url{https://data.gnss.ga.gov.au/docs/home/gnss-data.html}), and GDMS (\url{https://gdms.cwa.gov.tw/}). Ionosonde and HF Doppler sounding data were obtained from the Chinese Meridian Project (\url{https://www.meridianproject.ac.cn/}). Swarm L1B Electric Field Instrument (EFI) data are from the European Space Agency (ESA; \url{https://swarm-diss.eo.esa.int/}). COSMIC-2 Level 2 ionospheric products are from the University Corporation for Atmospheric Research (UCAR) COSMIC Data Analysis and Archive Center (CDAAC; \url{https://cdaac-www.cosmic.ucar.edu/}). TIMED/GUVI O/N$_2$ data are available from the TIMED/GUVI website (\url{http://guvitimed.jhuapl.edu/}). JPL GIM products were accessed via the NASA Crustal Dynamics Data Information System (\url{https://cddis.nasa.gov/archive/gnss/products/ionex/}). 
Ionosfera-M topside sounding figures included in the Supporting Information are available from the authors upon reasonable request.

\section*{Conflict of Interest}

The authors declare no conflicts of interest relevant to this study.

\bibliographystyle{plainnat}
\bibliography{references}

\end{document}